\documentclass[global,twocolumn]{svjour}

\usepackage{graphicx}
\usepackage{bm}

\journalname{Applied Physics B}

\begin{document}
\title{A Compact Microchip-Based Atomic Clock Based on Ultracold Trapped Rb Atoms}

\author{Daniel M. Farkas\inst{1} \and Alex Zozulya\inst{2} \and Dana Z. Anderson\inst{1}}

\institute{Department of Physics and JILA, University of Colorado, Boulder, CO 80309-0440, USA \and
Department of Physics, Worcester Polytechnic Institute, 100 Institute Road, Worcester, MA 01609, USA}

\date{Received: date / Revised version: date}

\maketitle

\begin{abstract}
We propose a compact atomic clock based on ultracold Rb atoms that are magnetically trapped near the surface of an atom microchip. An interrogation scheme that combines electromagnetically-induced transparency (EIT) with Ramsey's method of separated oscillatory fields can achieve atomic shot-noise level performance of $10^{-13}/\sqrt{\tau}$ for $10^6$ atoms. The EIT signal can be detected with a heterodyne technique that provides noiseless gain; with this technique the optical phase shift of a 100\,pW probe beam can be detected at the photon shot-noise level. Numerical calculations of the density matrix equations are used to identify realistic operating parameters at which AC Stark shifts are eliminated. By considering fluctuations in these parameters, we estimate that AC Stark shifts can be canceled to a level better than $2\times10^{-14}$. An overview of the apparatus is presented with estimates of duty cycle and power consumption.
\end{abstract}


\section{Introduction}
Recent progress in miniaturizing cold- and ultracold-atom systems has opened up the possibility of using ultracold matter and Bose-Einstein condensates (BECs) for portable applications. Inertial sensing, gravimetry, and timekeeping all stand to benefit from recent advances in the field~\cite{knemann_freely_2007,vogel_boseeinstein_2006,yu_development_2006,mcguirk_sensitive_2002,Stern2009,treutlein_coherence_2004,Ott2001,Hansel2001,Wang2005}. Some key developments include compact ultrahigh vacuum (UHV) systems, distributed-feedback diode lasers, and silicon fabrication techniques for both trapping with atom microchips and MEMS-based fiber-optic components. Several of these technologies were recently used to demonstrate BEC production in a robust, portable system having a volume of only 0.4\,m$^3$~\cite{Farkas2009}.

We propose a transportable, miniature atomic clock based on laser-cooled $^{87}$Rb atoms that are magnetically trapped near the surface of an atom microchip~\cite{treutlein_coherence_2004}.\linebreak Cooled, trapped atoms have significant potential in addressing needs for high-accuracy clocks having a small footprint. For example, Doppler shifts are significantly reduced by the cold temperatures, while the spatial confinement provided by the trap prevents decoherence from wall collisions. In addition, atom trapping can in principle make the clock insensitive to orientation in a gravitational field and can accommodate a dynamic environment subject to modest g-forces.  Moreover, implementation with an atom microchip enables one to consider its integration with other chip-based cold atom sensors.

Currently, chip-scale atomic clocks (CSACs) based on hot alkali vapors represent the state-of-the-art in \linebreak miniature atomic clocks~\cite{knappe_microfabricated_2004,vanier_practical_2005,vanier_atomic_2005}. Although instabilities down to 6$\times$10$^{-12}$ have been achieved, there are obstacles to achieving better performance from these devices. First, the coherence time of the atoms is shortened by collisions between the atoms and the cell walls. Although the use of buffer gases reduces dephasing, coherence \linebreak times are still limited to only a few ms. Second, this use of a buffer gas introduces pressure-dependent frequency shifts that are affected by temperature changes. Third, continuous excitation of the atoms by the laser field generates AC Stark shifts and linewidth broadening that contribute to long-term drift~\cite{v._shah_active_2006,zhu_theoretical_2000}.

With the limitations of clocks based on vapor cells on the one hand, and the unprecedented accuracies better than $10^{-16}$ in trapped, cold-atom macroscale clocks on the other
~\cite{rosenband_frequency_2008,ludlow_sr_2008}, cooling and trapping are advantageous towards enabling further progress in the development of miniature atomic clocks. Atom trapping is crucial for lengthening coherence times and hence improving clock stability. For example, consider a physics package measuring no more than 10\,cm in any dimension. Without trapping, the coherence time of an atom will be limited by the travel time of $\sim50\,$ms before gravity forces it to hit a wall, where the resulting collision will destroy the coherence. The resulting linewidth is $\sim$10\,Hz, regardless of how cold the atoms are. Permitting the clock's physical dimension to be even longer than 10\,cm in the direction of gravity, the atomic transverse temperature still limits travel times to less than 50\,ms, assuming typical atom temperatures from a magneto-optical trap ($\sim150\,\mu$K for $^{87}$Rb).

For magnetically trapped, ground state Rb atoms, coherence times greater than 2\,s have been observed experimentally~\cite{treutlein_coherence_2004,harber_effect_2002}. Our target for a background pressure of 10$^{-10}$\,Torr and atom temperature of less than 1$\,\mu$K should result in coherence times of at least 1\,s, providing an interrogation time equivalent to that of a 1.25\,m fountain clock. The resulting Ramsey fringe linewidth of 0.5\,Hz would be an order of magnitude narrower than what can be obtained with untrapped cold atoms and two orders of magnitude narrower than what can be obtained in a vapor cell. Such improvements more than compensate for the loss of atoms that accompany cooling and trapping.

Due to their insensitivity to magnetic fields, the \linebreak $m=0$ magnetic sublevels normally used as the clock states in alkali-atom clocks cannot be used. Instead, the low-field seeking, magnetically trappable $|F=1,m_F=-1\rangle$ and $|F=2,m_F=+2\rangle$ ground state sublevels can be used. In a magnetic field near 3.23\,G, the first-order Zeeman shift of this transition cancels, helping to reduce the sensitivity of the clock frequency to fluctuations and inhomogeneities in the trapping fields~\cite{treutlein_coherence_2004,harber_effect_2002}.

For interrogating the atoms, we consider Ramsey's method of separated oscillatory fields with pulsed electromagnetically induced transparency (EIT)~\cite{zanon-willette_cancellation_2006}. Implementations of Ramsey spectroscopy using pulsed EIT have been studied extensively in atomic beams and vapor cells~\cite{zanon_high_2005,thomas_observation_1982,hemmer_semiconductor_1993}. However, to our knowledge, no attempt has yet been made to experimentally observe this phenomenon in cold, trapped atoms, where it has been proposed as a way to eliminate AC Stark shifts in atomic clocks~\cite{zanon-willette_cancellation_2006,hemmer_semiconductor_1993,shahriar_dark-state-based_1997}.

This paper is divided into four sections. The first part is a basic overview of the proposed experimental approach. The second part presents theoretical calculations of Ramsey spectroscopy with pulsed EIT. From numerical solutions to the density matrix equations, clock stability is estimated for realistic operating parameters. The third part focuses on systematic effects. We show that for a proper choice of laser pulse durations and detunings, AC Stark shifts can be eliminated. Dephasing due to magnetic field inhomogeneity and mean-field shifts are also reviewed. In the last part, an overview of a proposed experimental implementation is presented with focus on a compact UHV vacuum system. Estimates of the duty cycle and power consumption for a first generation implementation are also presented.


\section{Experimental Overview}

\begin{figure}
\centering
\includegraphics[width=3.2in]{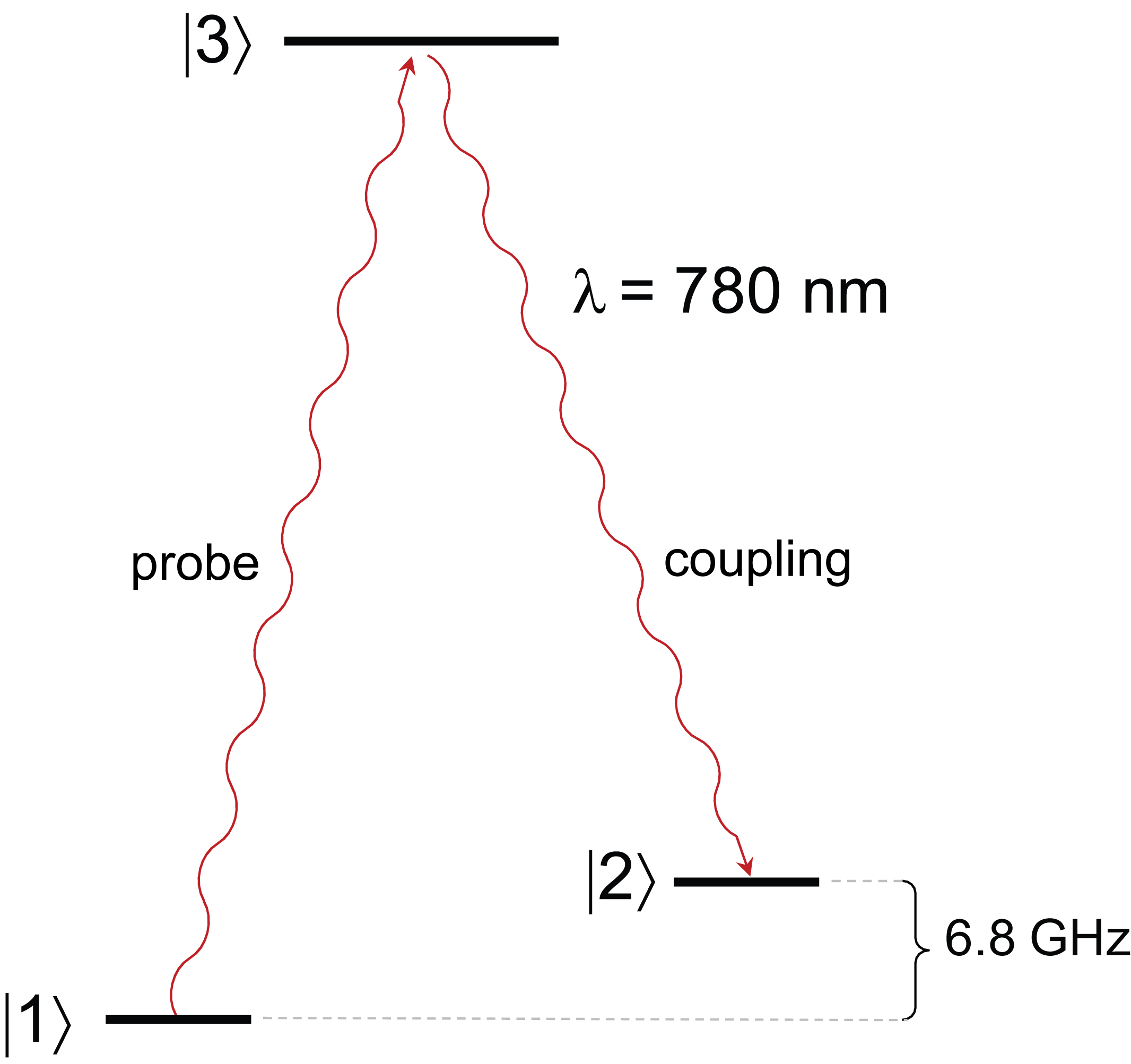}
\caption{Simplified energy diagram of $^{87}$Rb. The clock states $|1\rangle=|5^2S_{1/2},F=1,m_F=-1\rangle$ and $|2\rangle=|5^2S_{1/2},F=2,m_F=+1\rangle$ can be magnetically trapped and have an energy spacing equal to the $\sim6.8$\,GHz ground-state hyperfine splitting. The excited state $|3\rangle$ can be either the $|5^2P_{3/2},F'=2,m_{F'}=0\rangle$ or the $|5^2P_{3/2},F'=1,m_{F'}=0\rangle$ sublevel.}
\label{fig:RbEnergyDiagram}
\end{figure}
A simplified energy diagram of the $D_2$ transition in $^{87}$Rb is shown in Fig.~\ref{fig:RbEnergyDiagram}. The clock states $|1\rangle=|5^2S_{1/2},F=1,m_F=-1\rangle$ and $|2\rangle=|5^2S_{1/2},F=2,m_F=+1\rangle$ have an energy splitting $\omega_{21}$ equal to the $\sim$6.8\,GHz ground-state hyperfine splitting. These sublevels were chosen as clock states since they are both low-field seekers that can be trapped in a magnetic field minimum. A pair of laser beams with wavelengths near 780\,nm and frequencies separated by $\sim$6.8\,GHz are used to drive two electric dipole transitions: a $\sigma^+$-polarized probe beam drives the transition $|1\rangle \rightarrow |3\rangle$ and a $\sigma^-$-polarized coupling beam drives the transition $|2\rangle \rightarrow |3\rangle$. State $|3\rangle$ is one of the magnetic sublevels in the excited $5^2P_{3/2}$ manifold. Due to selection rules, this state must have magnetic component $m_{F'}=0$ and total angular momentum $F'=F \pm 1=1\,\,\mathrm{or}\,\,2$. Therefore the states $|5^2P_{3/2},F'=1,m_{F'}=0\rangle$ or $|5^2P_{3/2},F'=2,m_{F'}=0\rangle$ can serve as $|3\rangle$. Unprimed labels $F$ and $m$ refer to ground states in the $5^2S_{1/2}$ manifold, and primed labels are used for excited states in the $5^2P_{3/2}$ manifold.

\begin{figure}
\centering
\includegraphics[width=3.2in]{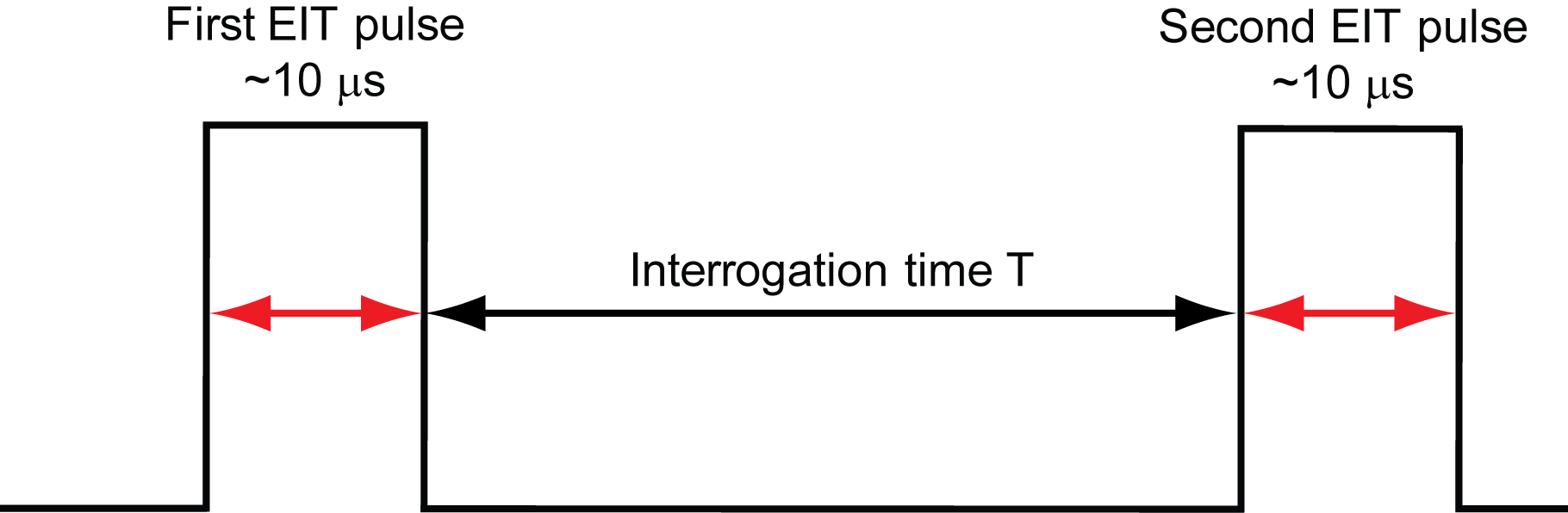}
\caption{The interrogation scheme uses pulsed EIT to implement Ramsey's method of separated oscillatory fields. The first pulse puts the atoms into a superposition of clock states. Atoms then precess in the dark for an interrogation time $T$, after which the laser difference frequency is compared to the clock transition frequency by measuring an optical phase shift inscribed on the probe beam during the second EIT pulse.}
\label{fig:EITpulses}
\end{figure}

All atoms must be in states $|1\rangle$ or $|2\rangle$ at the beginning of the operational sequence (see Fig.~\ref{fig:EITpulses}). This can be achieved either by optically pumping atoms into one of these states or simply waiting for atoms in other states, which are magnetically untrappable, to be ejected from the trapping region. Probe and coupling laser fields are then pulsed for duration $t_1$, using EIT to put the atoms into a steady-state, coherent superposition of states $|1\rangle$ and $|2\rangle$:
\begin{eqnarray}
\label{eqn:CoherentState}
| \psi \rangle = \frac{\Omega_c |1\rangle - \Omega_p |2\rangle}{\sqrt{\Omega_p^2+\Omega_c^2}},
\end{eqnarray}
where $\Omega_p$ and $\Omega_c$ are the Rabi frequencies of the probe and coupling transitions, respectively. We assume $\Omega_p = \Omega_c$, resulting in an equal superposition $(|1\rangle-|2\rangle)/\sqrt{2}$. In this case, the initial EIT pulse is functionally equivalent to the initial $\pi/2$ pulse in Ramsey spectroscopy.

After the laser fields are extinguished, the atoms free-ly precess in the dark for an interrogation time $T$. This step is equivalent to the second stage in Ramsey spectroscopy, yielding fringes of width $1/(2T)$\,Hz. The linear decrease in linewidth with interrogation time is one of the great benefits of Ramsey's method. By comparison, spectroscopy methods that continuously probe atoms, such as the coherent population trapping (CPT) interrogation done in CSACs, yield clock performances that improve only as the square-root of time.

The coherent superposition state created with the first EIT pulse is dark with respect to a pair of laser beams whose difference frequency is equal to $\omega_{21}$. This dark state is detected by applying a second EIT pulse of duration $t_2$ and monitoring the optical phase shift of the probe beam. If the laser difference frequency is equal to $\omega_{21}$, then the atomic sample remains transparent, and no phase shift is inscribed by the atoms onto the probe beam. If the laser difference frequency does not equal $\omega_{21}$, then the atoms will not be dark and the probe beam will pick up a phase shift. We assume that the Rabi frequencies of the second pulse are the same as those for the first pulse, otherwise the second pulse will pump atoms into a different superposition state, via Eqn.~\ref{eqn:CoherentState}, leading to a non-zero phase shift of the probe beam even if the laser difference frequency equals $\omega_{21}$.

For magnetically trapped Rb atoms, recent theoretical work has investigated the impact of field inhomogeneity and cold collisional shifts on coherence time, frequency drift, and accuracy \cite{rosenbusch_magnetically_2009}. We work within the parameters of references~\cite{harber_effect_2002} and~\cite{rosenbusch_magnetically_2009} and consider a sample of Rb atoms that has been evaporatively cooled to a temperature of $T_a=500$\,nK. Atoms are trapped near the surface of an atom microchip in a harmonic trap formed from electrical currents running through conductors on the chip surface. The atomic density will have a Gaussian spatial profile with a FWHM of $\sqrt{8\,\mathrm{ln}\,2\,k_B T_a/(m\,\omega^2)}$, where $k_B$ is Boltzmann's constant and $m$ is the atomic mass. For a transverse trap frequency of $\omega_t=100$\,Hz and an axial trap frequency of $\omega_a=10$\,Hz, the size of the atomic cloud is 25\,$\mu$m in the transverse directions and 250\,$\mu$m in the axial direction. For simplicity, we assume the atomic sample to have a uniform cylindrical density profile with these dimensions.

Mode-matching the waist of the probe beam to the very tight transverse dimension of the atomic cloud pre-sents a challenge for achieving shot-noise limited detection of the probe beam. To avoid saturation of the optical transitions (i.e. $\Omega_c$,$\Omega_p \ll (2\pi) 6$\,MHz), the probe power must be less than 1\,nW. Such low powers cannot be detected at the shot-noise level using conventional PIN or avalanche photodiodes. To help circumvent this limitation, we choose the $|F'=2,m_{F'}=0\rangle$ sublevel for the excited state $|3\rangle$. Compared to the $|F'=1,m_{F'}=0\rangle$ sublevel, the probe transition to this state is weaker, and therefore larger probe powers can be tolerated.

\begin{figure}
\centering
\includegraphics[width=3.4in]{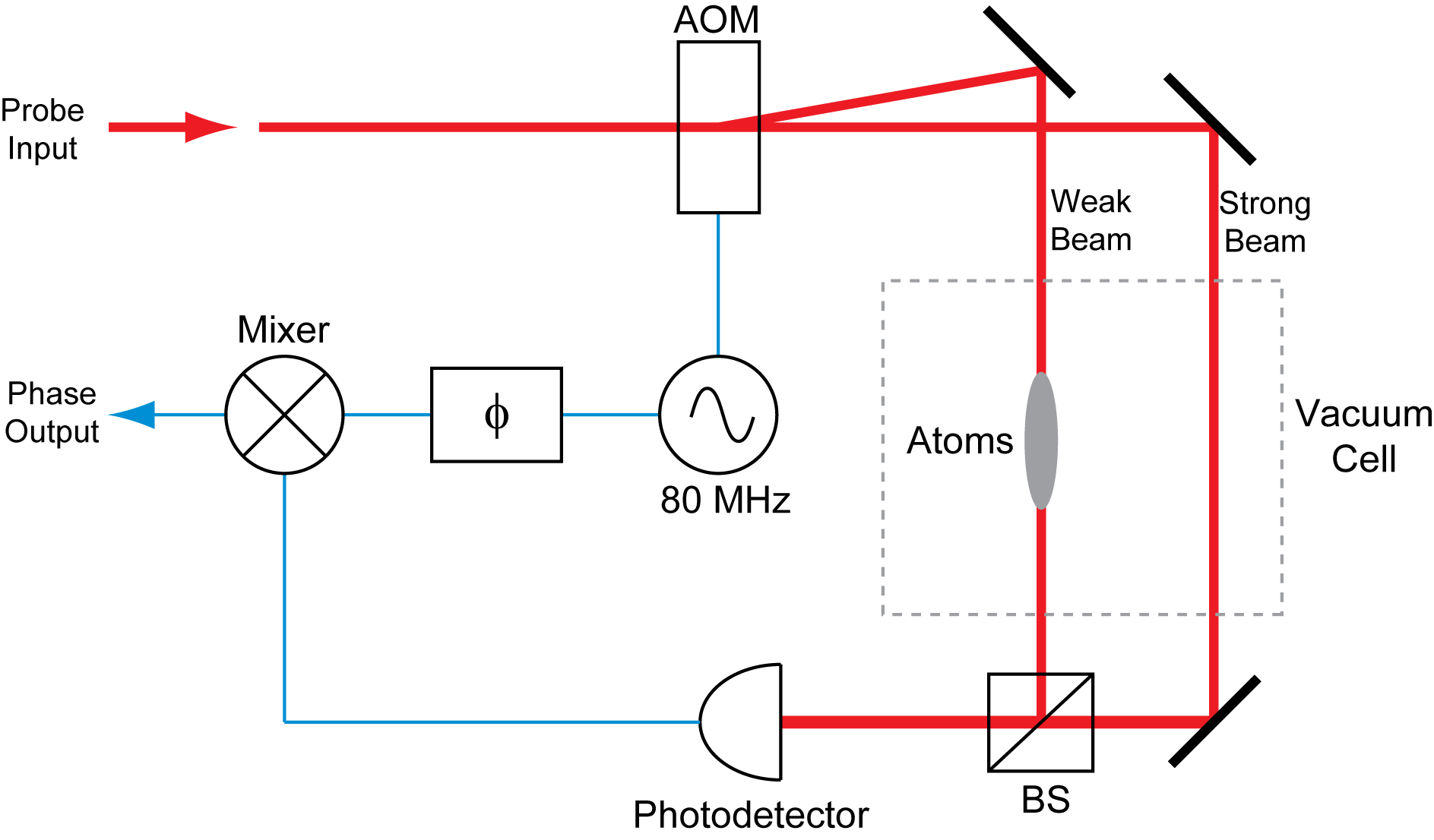}
\caption{Schematic for optical heterodyne detection of a 100\,pW beam at the shot-noise level. The probe is split into strong ($P_s\sim1\,$mW) and weak ($P_w\sim100\,$pW) beams. The weak beam is frequency shifted with an acousto-optic modulator (AOM) before passing through the atomic cloud. The weak and strong beam are recombined on the face of a photodetector. The electronic beat at the AOM frequency is demodulated using a phase-shifted copy of the AOM drive. The proper choice of phase $\phi$ selects between in-phase and quadrature signals, one of which is directly proportional to the phase shift inscribed by the atoms onto the weak beam.}
\label{fig:HeterodyneDetection}
\end{figure}
We present two optical-RF double heterodyne techniques for detecting laser powers as low as 100\,pW at the shot-noise level. In the first technique (see Fig.~\ref{fig:HeterodyneDetection}), the probe laser is split into weak and strong beams of powers $P_w$ and $P_s$, respectively. The weak beam ($P_w\sim100$\,pW) is shifted by $\sim$80\,MHz in an acousto-optic modulator (AOM) before passing through the atomic cloud. The strong beam ($P_s\sim1$\,mW) propagates parallel to the weak beam, but is laterally translated so that it does not interact with the atoms. The two beams are recombined with a beamsplitter and the heterodyne beat at 80\,MHz is detected with a silicon photodetector. The beat signal has an amplitude proportional to $\sqrt{P_w P_s}$, with the strong beam acting as a local oscillator that provides noiseless gain for detection of the weak beam. At 80\,MHz, technical noise on the lasers has rolled \linebreak off and the beat can be detected at the shot-noise limit. The shot noise is dominated by that of the strong beam and is proportional to $\sqrt{P_s}$. The resulting signal-to-noise ratio of $\sqrt{P_w P_s}/\sqrt{P_s}=\sqrt{P_w}$ is determined by the shot-noise of the weak beam. The electronic signal at 80\,MHz is demodulated with an adjustable phase $\phi$ that selects between the in-phase and quadrature components.

A limitation of this technique is that mechanical drift of the optics will introduce systematic errors. The success of this scheme, therefore, depends crucially on the mechanical stability of the setup. One method for avoiding this limitation is to repeat the probe measurement a second time without atoms. This second measurement calibrates the systematic error and can be used to correct the first measurement. The use of two measurements assumes that mechanical drift is slow compared to the time required to perform two probe measurements. For mechanical vibrations on faster timescales, a dual-wavelength interferometer can be used. Here, systematic phase shifts are measured at the same time as the atom-induced phase shift using a laser beam at a wavelength much different than that of the probe.

The second heterodyne detection technique is based on phase modulation of the probe beam. Here, strong modulation generates a weak carrier with strong sidebands that provide noiseless gain for carrier detection. In the absence of an atom-induced optical phase shift, a phase modulated beam will produce no electronic beat signal upon photodetection. However, once the carrier is phase shifted, the phase modulation is partially converted into amplitude modulation that can be electronically detected. The benefit of this approach is that optical phase shifts due to mechanical instabilities affect all frequency components identically; this common-mode systematic noise will therefore be rejected upon detection. However, the drawback to this approach is the rise of additional AC Stark shifts due to the multiple frequency components of the probe beam.


\section{Theoretical Calculations}

In order to estimate clock stability, we numerically solved the density matrix equations for the three-level system. The solutions are used to obtain the susceptibility, from which the optical phase shift inscribed by the atoms onto the probe beam can be calculated. The phase shift is numerically integrated as a function of time to obtain theoretical Ramsey fringes as well as numerical estimates for signal-to-noise ratio and clock stability.

\subsection{The Dynamical Equations of Motion}
Consider the three-level system shown in Fig.~\ref{fig:RbEnergyDiagram} in the presence of an electromagnetic field given by
\begin{eqnarray}\label{general_field}
    \hspace{-0.2in} \bm{\mathcal{E}} = \frac{1}{2}\left[\bm{A}_{31}\exp(-i\omega_{31}t)+\bm{A}_{32}\exp(-i\omega_{32}t) + \mathrm{c.c.}\right].
\end{eqnarray}
The field consists of two components: a probe beam of amplitude $\bm{A}_{31}$ and frequency $\omega_{31}$ that is nearly resonant with the transition $|1\rangle\rightarrow|3\rangle$, and a coupling beam of amplitude $\bm{A}_{32}$ and frequency $\omega_{32}$ that is nearly resonant with the transition $|2\rangle\rightarrow|3\rangle$. Light changes the optical properties of the medium because it affects the evolution of the atomic density operator $\rho$. The evolution equation for $\rho$ is
\begin{equation}\label{general_rho}
    \frac{d}{dt}\rho  = -\frac{i}{\hbar}[H_{a} + H_{int},\rho] + R_{\rho},
\end{equation}
where $H_{a}$ is the atomic Hamiltonian in the absence of light and $H_{int} = -\bm{D} \cdot {\bm{\mathcal E}}(t)$ is the dipole interaction Hamiltonian between the light field and the atom. $\bm{D}$ is the dipole moment operator and $R_{\rho}$ is the relaxation operator.

An electric dipole transition between clock states $|1\rangle$ and $|2\rangle$ is not allowed by selection rules. The resulting atomic dipole operator has the form
\begin{equation}\label{general_dipole}
    {\bm D} = |3\rangle \langle 1|{\bm \mu_{31}} + |3\rangle \langle 2|{\bm \mu_{32}} + \mathrm{c.c.},
\end{equation}
where $\bm{\mu}_{mn} = \langle m|{\bm D}|n\rangle$ is the dipole transition matrix element. The Hamiltonian $H = H_{a} + H_{int}$ in the rotating-frame approximation can be written
\begin{eqnarray}\label{general_H}
    && \hspace{-0.25in} H = \hbar \omega_{1}|1\rangle \langle 1| + \hbar \omega_{2}|2\rangle \langle 2| + \hbar \omega_{3}|3\rangle \langle 3| \nonumber \\
    && \hspace{-0.25in} {}-\frac{\hbar}{2}\left(\Omega_{31}|3\rangle \langle 1|e^{-i\omega_{31}t} + \Omega_{32}|3\rangle \langle 2|e^{-i\omega_{32}t} +\mathrm{c.c.} \right)
\end{eqnarray}
where $\Omega_{31} = \bm{\mu}_{31}\cdot \bm{A}_{31}/\hbar$ and $\Omega_{32} = \bm{\mu}_{32}\cdot \bm{A}_{32}/\hbar$ are the Rabi
frequencies of the probe and coupling transitions, respectively.

Evolution of the electromagnetic field is governed by Maxwell's equations, which reduce to
\begin{equation}\label{general_maxwell}
    \nabla^{2}\bm{\mathcal{E}} -
    \frac{1}{c^{2}}\frac{\partial^{2}}{\partial t^2}\left(\bm{\mathcal{E}} +
    \frac{1}{\epsilon_{0}}\bm{\mathcal{P}}\right) = 0,
\end{equation}
where $\bm{\mathcal{P}}$ is the polarization of the medium. The atomic dipole operator is given by Eqn.~\ref{general_dipole}, so the single atom polarization is
\begin{equation}\label{general_1atom_polarization}
    {\bm p} = {\bm \mu}_{13}\rho_{31} + {\bm \mu}_{23}\rho_{32} + \mathrm{c.c}.
\end{equation}
The polarization $\bm{\mathcal{P}}$ of the medium is given by the relation
\begin{eqnarray}\label{general_polarization}
    \bm{\mathcal{P}} &=& \frac{1}{2}(\bm{P}_{31} + \bm{P}_{32} + \mathrm{c.c.}) \\
    &=& n_{a}(\bm{\mu}_{13}\rho_{31} + \bm{\mu}_{23}\rho_{32} + \mathrm{c.c.})
\end{eqnarray}
where $n_{a}$ is the density of atoms. The steady-state response of the term ${\bm P}_{31}$ oscillates at the frequency $\omega_{31}$ $({\bm P}_{31} \propto \exp(-\omega_{31}t))$
and the term ${\bm P}_{32}$ oscillates at the frequency $\omega_{32}$.

Eqns.~\ref{general_rho} and \ref{general_H} show that the coherences, or off-diagon-al elements of the density matrix, can be written as $\rho_{31} = \sigma_{31}\exp(-i\omega_{31}t)$ and $\rho_{32} = \sigma_{32}\exp(-i\omega_{32}t)$, where the functions $\sigma_{ij}$ are slow functions of time as compared with the exponentials. It is convenient to factor out the fast time dependence from the problem and work with slow-varying elements of the density matrix. This is achieved by going to the interaction representation, i.e. factorizing the Hamiltonian in Eqn.~\ref{general_H} as $H = H_{0} + V$ and introducing the interaction-representation density operator $\sigma$ by the relation
$\sigma = \exp(iH_{0}t/\hbar)\rho \exp(-iH_{0}t/\hbar)$.
The density operator $\sigma$ obeys the equation
\begin{equation}\label{interaction_pic1}
    \frac{d}{dt}\sigma = -\frac{i}{\hbar}[V_{I},\sigma] + R_{\sigma},
\end{equation}
where
\begin{equation}\label{VI_definition}
    V_{I} = \exp(iH_{0}t/\hbar)V\exp(-iH_{0}t/\hbar)
\end{equation}
is the interaction Hamiltonian. Similarly, $R_{\sigma}$ is the relaxation operator in the interaction representation.

The Hamiltonian $H_{0}$ should be chosen so that elements of the density matrix $\sigma$ are slow functions of time, i.e. do not contain fast time scales on the order of inverses of laser frequencies. A convenient choice is
\begin{eqnarray}\label{H0}
    && \hspace{-0.6in} H_{0} = \hbar(\omega_{3} - \omega_{31})|1\rangle \langle 1| \nonumber \\
    && {}+ \hbar (\omega_{3} - \omega_{32})|2\rangle \langle 2| + \hbar \omega_{3}|3\rangle \langle 3|
\end{eqnarray}
and
\begin{eqnarray}
    && \hspace{-0.25in} V = \hbar\Delta_{31}|1\rangle \langle 1| + \hbar \Delta_{32}|2\rangle \langle 2| \nonumber \\
    && \hspace{-0.3in} {}-\frac{\hbar}{2}\left(\Omega_{31}|3\rangle \langle 1|e^{-i\omega_{31}t} +  \Omega_{32}|3\rangle \langle 2|e^{-i\omega_{32}t} + \mathrm{c.c.}\right),
\end{eqnarray}
where $\Delta_{31} = \omega_{31} - (\omega_{3} - \omega_{1})$ and $\Delta_{32} = \omega_{32} - (\omega_{3} - \omega_{2})$ are the detunings of the
two lasers from the frequencies of the corresponding transitions with which they are nearly resonant.

The slowly-varying density matrix $\sigma$ is related to the full density matrix $\rho$ by
\begin{equation}\label{rho_and_sigma}
    \hspace{-0.1in} \sigma = \left[\begin{array}{ccc}
    \rho_{11} & \rho_{12}e^{-i(\omega_{31} - \omega_{32})t} & \rho_{13}e^{-i\omega_{31}t} \\
    \rho_{21}e^{i(\omega_{31} - \omega_{32})t} & \rho_{22} & \rho_{23}e^{-i\omega_{32}t} \\
    \rho_{31}e^{i\omega_{31}t} & \rho_{32}e^{i\omega_{32}t} & \rho_{33}
    \end{array}\right].
\end{equation}
The interaction Hamiltonian $V_{I}$ in Eqn.~\ref{VI_definition} is given by the expression
\begin{equation}\label{VI}
    V_{I} = \hbar \left[\begin{array}{ccc}
    \Delta_{31} & 0 & - \Omega_{13}/2 \\
    0 & \Delta_{32} & - \Omega_{23}/2 \\
    - \Omega_{31}/2 & - \Omega_{32}/2 & 0
    \end{array}\right].
\end{equation}
The relaxation matrix is given by
\begin{equation}
    R\sigma = \left[\begin{array}{ccc}
    \Gamma_{31}\sigma_{33} & -\gamma_{c}\sigma_{12} & -\gamma_{c1}\sigma_{13} \\
    -\gamma_{c}\sigma_{21} & \Gamma_{32}\sigma_{33} & - \gamma_{c2}\sigma_{23} \\
    -\gamma_{c1}\sigma_{31} & - \gamma_{c2}\sigma_{32} & - \Gamma\sigma_{33}
    \end{array}\right].
\end{equation}
The relaxation matrix includes the total spontaneous decay rate $\Gamma$ of the excited state; partial decay rates $\Gamma_{31}$ and $\Gamma_{32}$ of the excited state to the clock states $|1\rangle$ and $|2\rangle$, respectively; optical decoherences $\gamma_{c1}$ and $\gamma_{c2}$; and the Raman decoherence $\gamma_{c}$. The system is assumed to be open, implying that atoms are lost upon spontaneously decaying to states other than $|1\rangle$ or $|2\rangle$. As a result, the total decay rate of the excited state $\Gamma=(2\pi)6.0$\,MHz is larger than the sum of the partial decay rates (see Table \ref{tab:RbProperties}). We ignore decoherence between the clock states (i.e. $\gamma_{c} = 0$) and set the optical decoherences equal to half of their corresponding partial decay rates (i.e. $\gamma_{c1}=\Gamma_{31}/2$ and $\gamma_{c2}=\Gamma_{32}/2$).

The set of equations for the elements of the density matrix takes the form
\begin{eqnarray}\label{3level_eqs_4_sigma}
    \frac{d}{dt}\sigma_{21} &=& (i\delta - \gamma_{c})\sigma_{21}
    + \frac{i}{2}[\Omega_{23}\sigma_{31} - \Omega_{31}\sigma_{23}], \nonumber \\
    \frac{d}{dt}\sigma_{31} &=& \left[i(\Delta_0+\frac{\delta}{2}) - \gamma_{c1}\right]\sigma_{31} \nonumber \\
    && + \frac{i}{2}[\Omega_{31}(\sigma_{11} - \sigma_{33}) +
    \Omega_{32}\sigma_{21}], \nonumber \\
    \frac{d}{dt}\sigma_{23} &=& \left[-i(\Delta_0-\frac{\delta}{2}) - \gamma_{c2}\right]\sigma_{23} \\
    && - \frac{i}{2}[\Omega_{23}(\sigma_{22} - \sigma_{33}) +
    \Omega_{13}\sigma_{21}], \nonumber \\
    \frac{d}{dt}\sigma_{11} &=&  \Gamma_{31}\sigma_{33} + \frac{i}{2}[\Omega_{13}\sigma_{31}
     - \rm{c.c.}], \nonumber \\
    \frac{d}{dt}\sigma_{22} &=& \Gamma_{32}\sigma_{33} - \frac{i}{2}[\Omega_{32}\sigma_{23}
     - \rm{c.c.}], \nonumber \\
    \frac{d}{dt}\sigma_{33} &=& - \Gamma \sigma_{33}  + \frac{i}{2}[- \Omega_{13}\sigma_{31}+\Omega_{32}\sigma_{23} - \mathrm{c.c.}], \nonumber
\end{eqnarray}
where the Raman detuning $\delta = \Delta_{31} - \Delta_{32} = (\omega_{31} - \omega_{32}) - (\omega_2-\omega_1)$ is the difference between laser difference frequency and the atomic clock frequency $\omega_{21}=\omega_2-\omega_1$. The common-mode detuning $\Delta_0=(\Delta_{31}+\Delta_{32})/2$ is the average of the two laser detunings.

The polarization of the atomic sample is given by
\begin{eqnarray}\label{polarization_sigma}
    \bm{\mathcal{P}} &=& \frac{1}{2}(\bm{P}_{31} + \bm{P}_{32} + \mathrm{c.c.}) \nonumber \\
    &=& n_{a}[\bm{\mu}_{13}\sigma_{31}e^{-i\omega_{31}t} + \bm{\mu}_{23}\sigma_{32}e^{-i\omega_{32}t} + \mathrm{c.c.}].
\end{eqnarray}
The relative dielectric constant of the medium is $\epsilon = 1 + \chi$, where $\chi$ is the medium susceptibility. Using Eqns.~\ref{general_maxwell} and~\ref{polarization_sigma}, $\chi$ becomes
\begin{eqnarray}\label{two_chis}
    \chi(\omega_p) = \frac{2n_{a}\Omega_{13}\hbar\sigma_{31}}{\epsilon_{0}|\bm{A}_{p}|^{2}}, \nonumber \\
    \chi(\omega_c) = \frac{2n_{a}\Omega_{23}\hbar\sigma_{32}}{\epsilon_{0}|\bm{A}_{c}|^{2}},
\end{eqnarray}
where $\omega_p$ and $\omega_c$ are the probe and coupling optical frequencies, respectively.

\subsection{Numerical Analysis}
\begin{table}
\centering
\begin{tabular}{l|cc}
\hline\noalign{\smallskip}
& Probe & Coupling \\
\noalign{\smallskip}\hline\noalign{\smallskip}
Transition & $|1\rangle \rightarrow |3\rangle$ &  $|2\rangle \rightarrow |3\rangle$\\
Partial Decay Rate (MHz) &  0.51 & 1.52 \\
Dipole Moment (C$\times$m) &  $7.32\times10^{-30}$ & $1.27\times10^{-29}$\\
Rabi Frequency (kHz) & 150 & 150\\
Intensity (W/m$^2$) & 0.245 & 0.082 \\
Power (pW) & 128 & 43\\
\noalign{\smallskip}\hline
\end{tabular}
\caption{Transition properties used for numerical simulations~\cite{steck_rubidium_2009}. The excited magnetic sublevel $|F'=2,m_{F'}=0\rangle$ is used for state $|3\rangle$. The intensities of the two lasers are chosen so that the Rabi frequencies for the two transitions are the same. For obtaining laser power from intensity, a beam waist of 25\,$\mu$m is assumed.}
\label{tab:RbProperties}
\end{table}
Eqns.~\ref{3level_eqs_4_sigma} were solved numerically using properties for the relevant transitions in $^{87}$Rb (see Table~\ref{tab:RbProperties}). Both probe and coupling lasers are on resonance with their respective transitions (i.e. $\delta=0$ and $\Delta_0=0$). Several factors must be considered when choosing appropriate values for the Rabi frequencies. A longer response time, as needed for a smaller Rabi frequencies, permits the electronic signal processing to operate at lower bandwidths, which simplifies the detection circuitry. On the other hand, smaller Rabi frequencies result in lower optical powers that in turn leads to lower detection signal-to-noise ratios. Taking these considerations into account, we chose $\Omega_c=\Omega_p=(2\pi)150$\,kHz for numerical estimates. Assuming a beam waist of 25\,$\mu$m, the corresponding probe power is 130\,pW.

\begin{figure}
\centering
\includegraphics[width=3.3in]{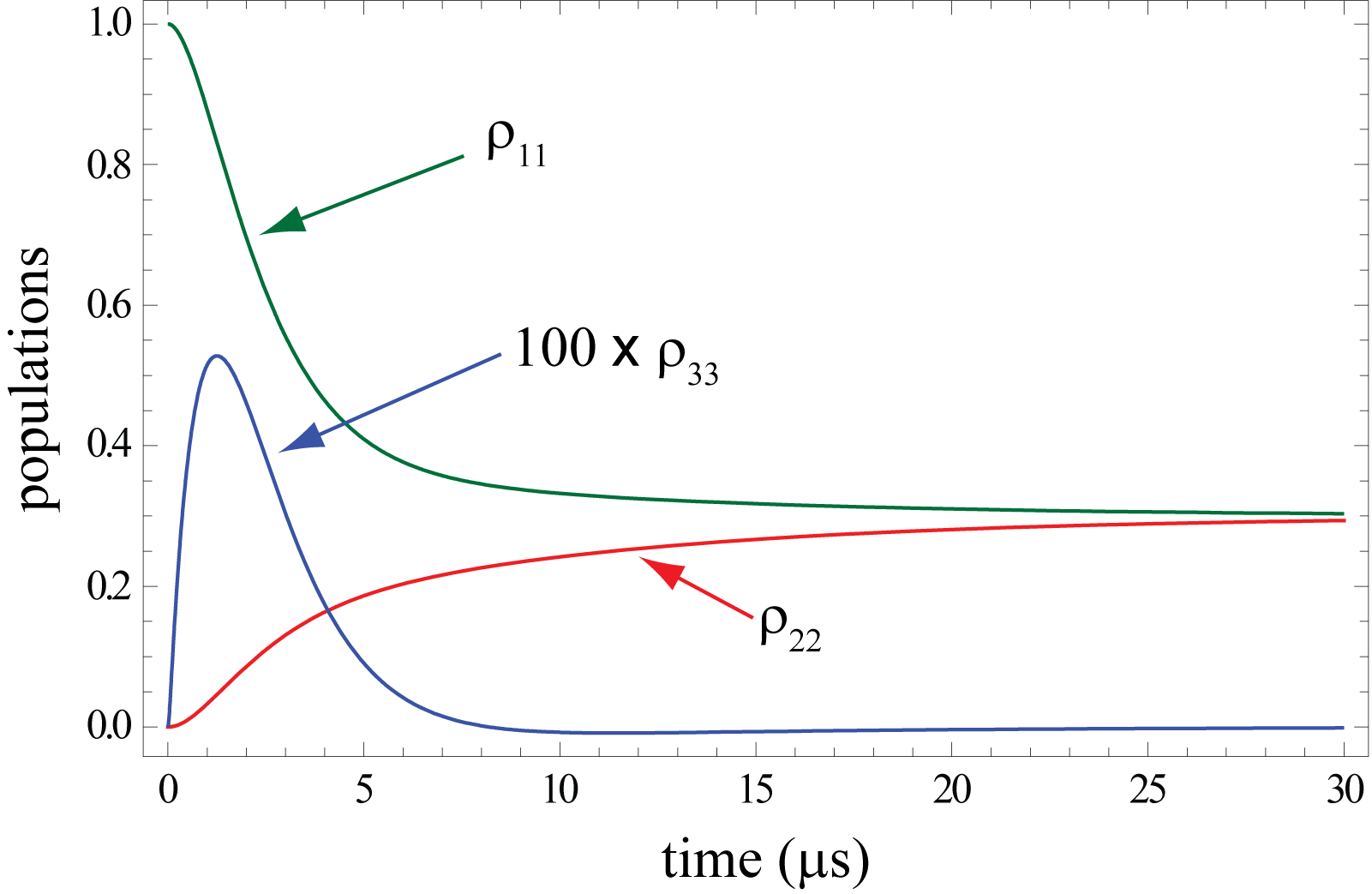}
\caption{The change in atomic populations as the initial EIT pulse is applied. The atoms start in the atomic eigenstate $|1\rangle$ ($\rho_{11}=1$). For $\Omega_{p}=\Omega_{c}$, the steady-state populations of the ground states are equal ($\rho_{11}=\rho_{22}$). The excited state becomes briefly populated ($\rho_{33}>0$) before attaining its steady-state value of 0.}
\label{fig:InitialPulsePopulations}
\end{figure}

Fig.~\ref{fig:InitialPulsePopulations} illustrates how the probe and coupling beams modify the density matrix populations. We assume the atomic sample starts entirely in state $|1\rangle$ (i.e. $\rho_{11}=1$,~$\rho_{22}=\rho_{33}=0$), although any choice of initial conditions will lead to the same steady-state response of equal clock state populations ($\rho_{11}=\rho_{22}$). The steady-state population $\rho_{33}=0$ in the presence of resonant laser fields is the clearest indication that the sample becomes optically transparent. Note that the excited state population $\rho_{33}$ briefly becomes non-zero. Although a large fraction of this excited state population decays into the dark coherent state, the remainder decays to ground state sublevels other than the clock states. Atoms in these other ground states will no longer be trapped and can therefore escape from the atomic cloud. The result is steady-state populations that sum to less than 1. Assuming these untrapped atoms are initially at rest, either gravity or a magnetic force will force them to fall out of the trap in less than 5 ms (ignoring collisions with other atoms). Although these untrapped atoms remain in the atomic cloud for the entire duration of the first pulse, they are not resonant with probe and coupling beams due to the magnetic field that lifts the degeneracy of magnetic sublevels.

\begin{figure}
\centering
\includegraphics[width=3.3in]{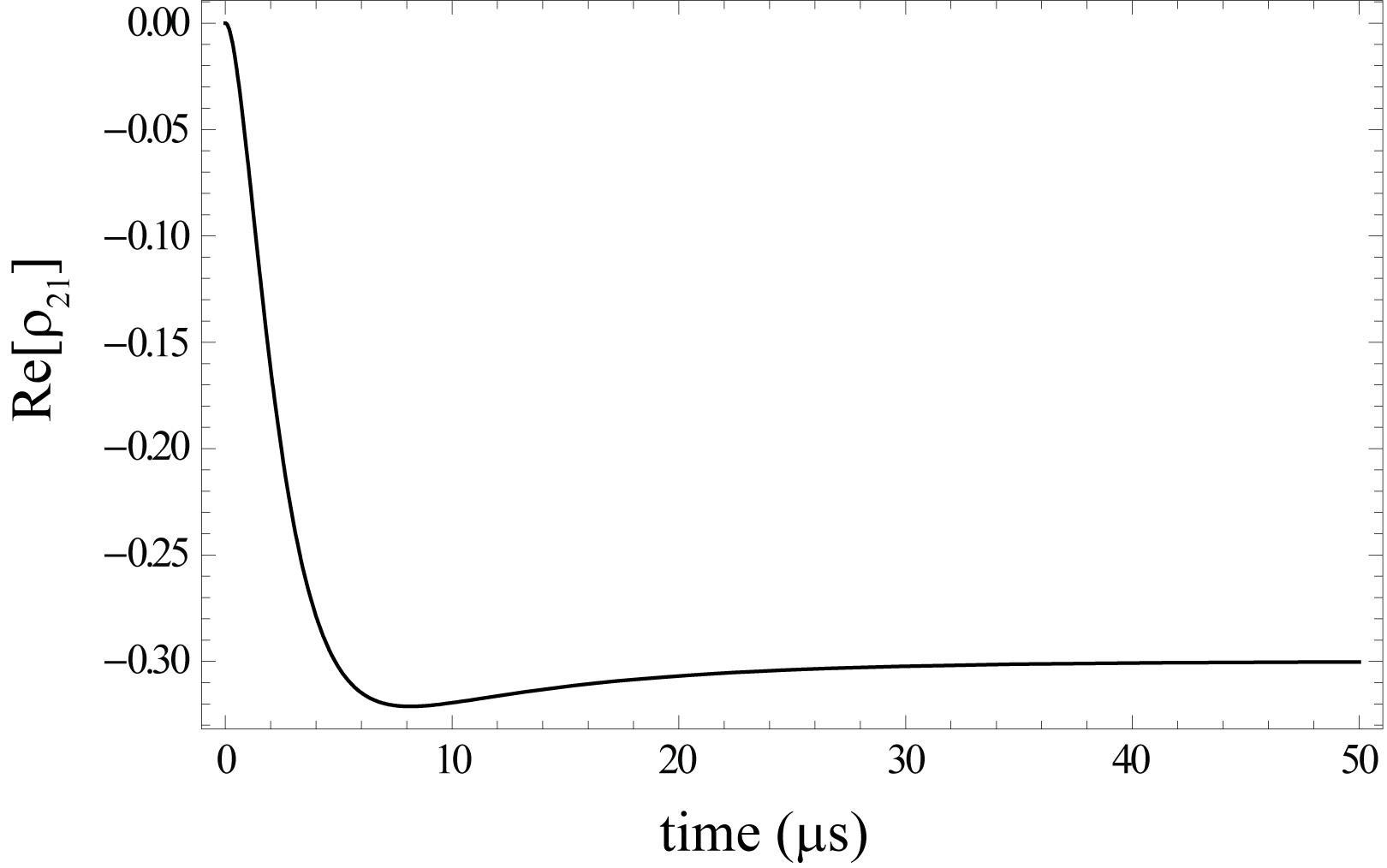}
\caption{The change in ground state coherence between the two clock states during the initial EIT pulse. Pumping of the atoms into a coherent superposition state gives rise to a non-zero, steady-state value of $\rho^{(1)}_{21}$.}
\label{fig:InitialPulseCoherence}
\end{figure}
Fig.~\ref{fig:InitialPulseCoherence} shows how coherence between the clock states is created during the initial pulse. For an equal superposition state, the coherence between clock states, given by the off-diagonal element $\rho_{21}$ of the density matrix, attains a final value $\rho^{(1)}_{21}$ equal in magnitude to the clock state populations $\rho_{11}$ and $\rho_{22}$. It is this non-zero coherence that differentiates atoms in a superposition state from atoms in equal populations of two different atomic eigenstates.

After the initial EIT pulse, atoms precess in the dark for time $T$, during which the coherence $\rho^{(1)}_{21}$ picks up a phase shift factor $\mathrm{exp}(-\delta\,T)$. After precession, the second EIT pulse is applied. We assume that values of the Rabi frequencies $\Omega_c$ and $\Omega_p$, Raman detuning $\delta$, and common-mode detuning $\Delta_0$ are equal to those used for the first pulse. Eqns.~\ref{3level_eqs_4_sigma} are again solved numerically, this time using $\rho^{(1)}_{21} \mathrm{exp}(-\delta\,T)$ for the initial value of the coherence. The clock state populations obtained at the end of the first pulse are used as the initial values for the second pulse.

\begin{figure}
\centering
\includegraphics[width=3.3in]{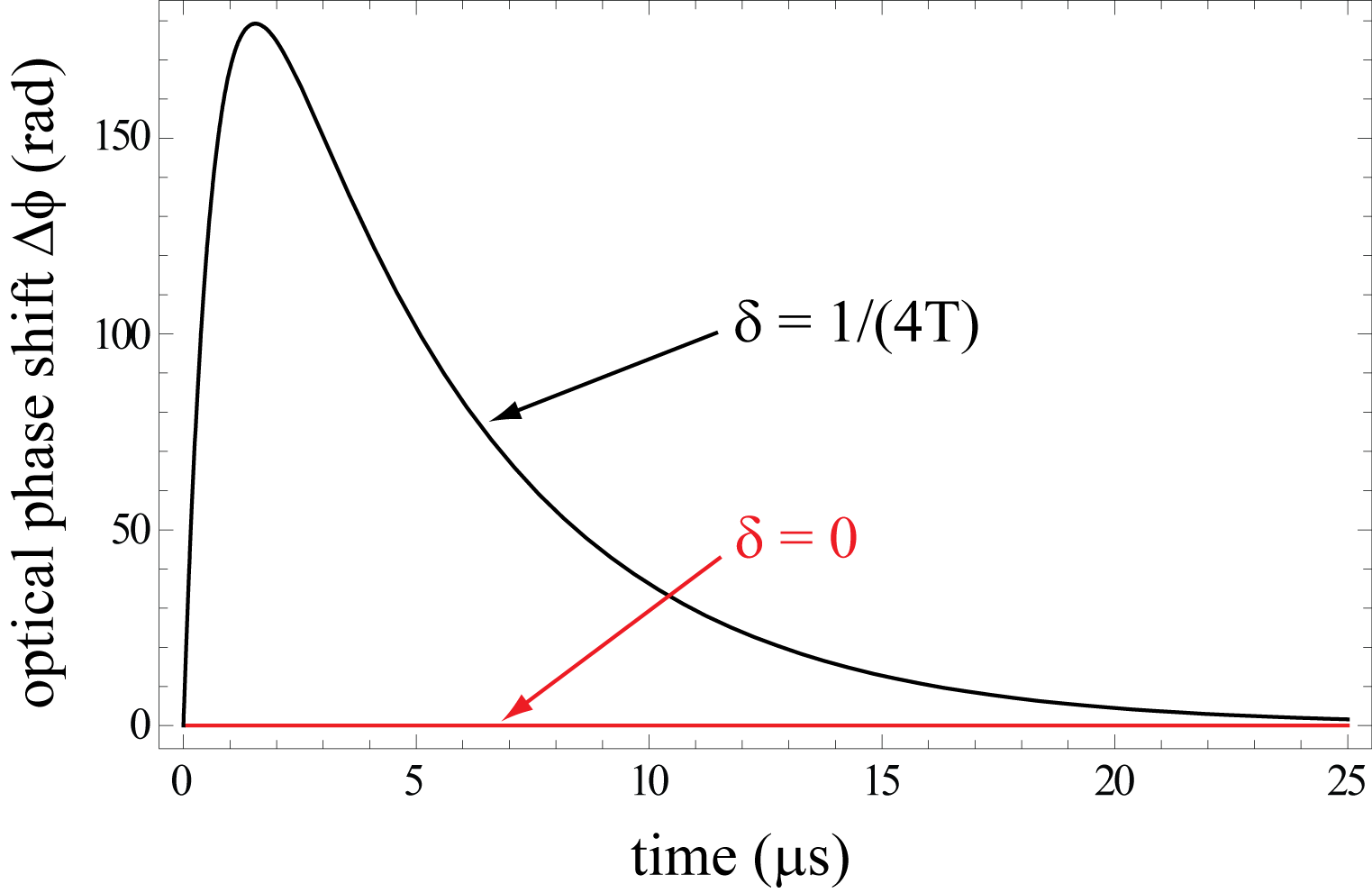}
\caption{Optical phase shift as a function of time during the second EIT pulse. When the Raman detuning is on resonance ($\delta=0$), the atomic sample is transparent and the probe beam picks up no phase shift from the atoms. The largest phase shifts occur for $\delta=(n/2+1/4)/T$, where $n$ is an integer.}
\label{fig:OpticalPhaseShift}
\end{figure}
The optical phase shift is given by $\Delta\phi = 2 \pi l (n'-1) / \lambda$, where $l$ is the length of the atomic sample in the propagation direction of the probe laser and the real component of the refractive index is given by \linebreak $n'=\mathrm{Re} [ \sqrt{1+\chi (\omega_p)} ]$. Fig.~\ref{fig:OpticalPhaseShift} illustrates the calculated optical phase shift as a function of time. When the laser difference frequency is resonant with the clock frequency $\omega_{21}$ (i.e. $\delta=0$), the sample is transparent and there is no phase shift (red). The maximum phase shift (black) occurs for $\delta = \pm 2\pi \times 1/(4T)$\,Hz.

\begin{figure}
\centering
\includegraphics[width=3.3in]{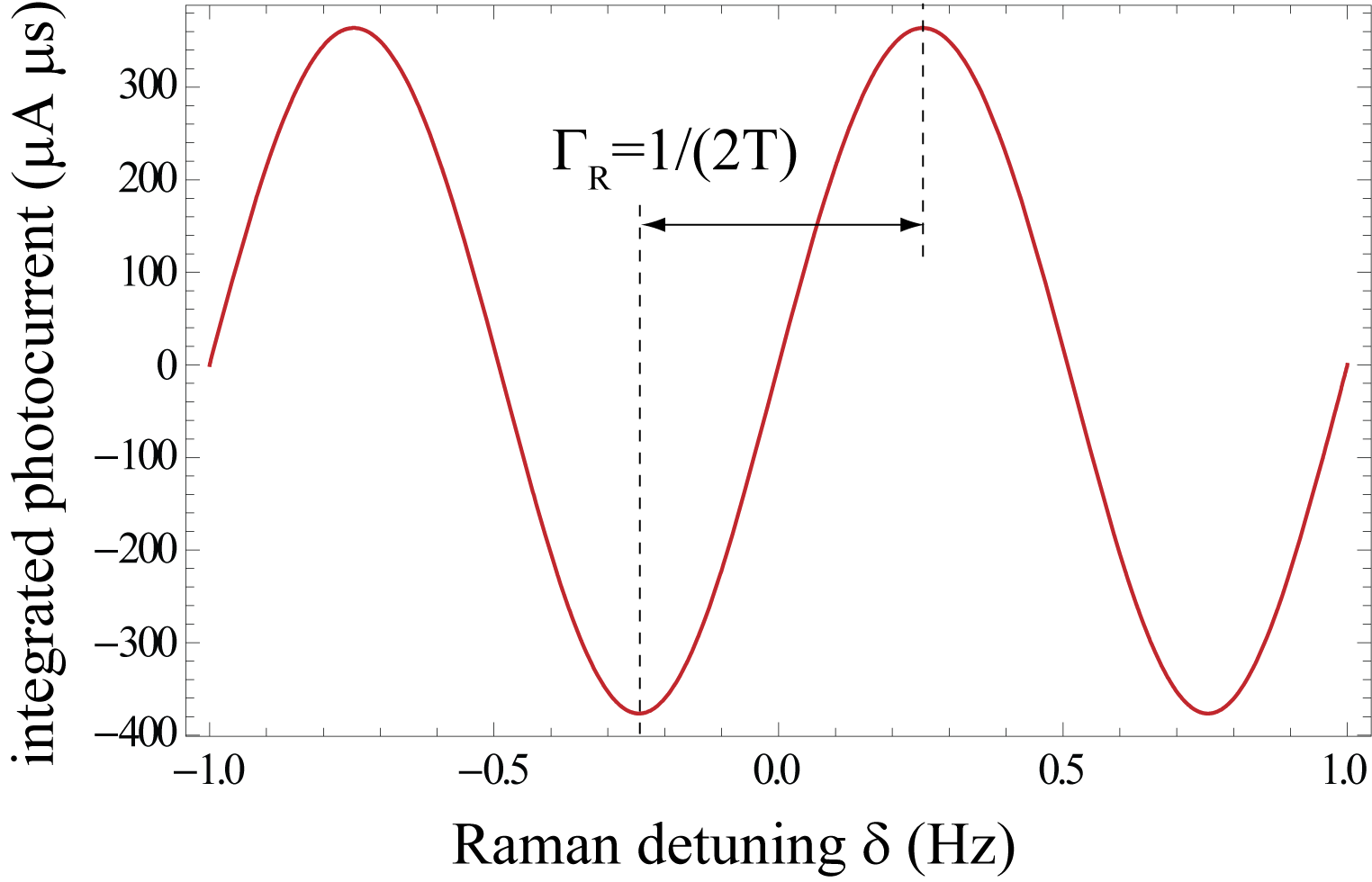}
\caption{Integrated optical phase shift as a function of Raman detuning $\delta$ shows Ramsey fringes with a FWHM of $\Gamma_R=1/(2T)$, where $T$ is the interrogation time.}
\label{fig:RamseyFringes}
\end{figure}
The detected photocurrent is given by \linebreak $I = 2 \mathcal{R} \sqrt{P_w P_s} \phi$, where $\mathcal{R}$ is the responsivity of the photodetector. To maximize signal-to-noise, the photocurrent is numerically integrated up to time $t_2$. Ramsey fringes of FWHM $\Gamma_R=1/(2T)$ are visible in Fig.~\ref{fig:RamseyFringes}, which shows the integrated photocurrent as a function of Raman detuning $\delta$. Unlike fringes that are obtained from absorption measurements, fringes obtained from optical phase shifts are dispersive and therefore have a zero-crossing at $\delta=0$. As a result, for small $\delta$, the integrated photocurrent is directly proportional to the correction that must be applied to the laser difference frequency.

\subsection{Signal-to-Noise Ratio (SNR) and Frequency Instability}
\label{sec:SNR}
The expected noise is obtained by integrating the noise current $i_n$ over the duration $t_2$ of the second EIT pulse. This yields a noise charge $q_n$. Assuming the photodetection noise is determined by the photon shot noise of the strong probe laser, the rms noise current $\sigma_i$ is given by
\begin{eqnarray}
\sigma_i = \sqrt{\langle i^2_n \rangle - \langle i_n \rangle^2} = \sqrt{2 e \mathcal{R} P_s \Delta f},
\end{eqnarray}
where $e$ is the electron charge and the circuit bandwidth $2 \pi \Delta f = 1/ \tau_R$ is the inverse of the characteristic response time $\tau_R$ of the circuit. The noise charge $q_n$ exhibits a random walk of duration $t_2$, yielding a mean-square noise charge $\langle q_n^2 \rangle$ of
\begin{eqnarray}
\label{eqn:Diffusion}
\langle  q_n^2 \rangle = 2\,D\,t_2,
\end{eqnarray}
where $D = \sigma_i\,\tau_R$ is the diffusion coefficient. Eqn.~\ref{eqn:Diffusion} assumes that the circuit response time $\tau_R$ is much faster than $t_2$ (i.e. $\tau_R \ll t_2$). The rms noise charge $\sigma_q$ can then be expressed
\begin{eqnarray}
\sigma_q = \sqrt{\langle q_n^2 \rangle - \langle q_n \rangle^2} = \sqrt{2 e \mathcal{R} P_s t_2 / \pi}.
\end{eqnarray}
Note that in the limit $\tau_R \ll t_2$, $\sigma_q$ depends only on $t_2$ and not on the circuit bandwidth.

\begin{table}
\centering
\begin{tabular}{l|c}
\hline\noalign{\smallskip}
Parameter & Value \\
\noalign{\smallskip}\hline\noalign{\smallskip}
Atom Number                             & $10^5-10^6$                   \\
Atom Temperature $T_a$                  & 500 nK                        \\
Atom Cloud Length $l$                   & 250 $\mu$m                    \\
Atom Cloud Waist                        & 25 $\mu$m                     \\
Peak Atom Density $n_a$                 & $10^{13} - 10^{14}$ cm$^{-3}$ \\
First Pulse Duration $t_1$              & $\sim$10 $\mu$s               \\
Interrogation Time $T$                  & 1 s                           \\
Fringe FWHM $\Gamma_R$                  & 0.5 Hz                        \\
Second Pulse Duration $t_2$             & $\sim$10 $\mu$s               \\
Photodiode Responsivity $\mathcal{R}$   & 0.5 A/W at 780 nm             \\
Weak Probe Power $P_w$                  & 130 pW                        \\
Strong Probe Power $P_s$                & 1 mW                          \\
\noalign{\smallskip}\hline
\end{tabular}
\caption{Assumed values for operational parameters.}
\label{tab:OperatingParams}
\end{table}

The signal is obtained from the slope $S$ of the Ramsey fringe near $\delta=0$. The signal-to-noise ratio SNR=$S$/$\sigma_q$ is numerically calculated to be $\sim$7.5$\times 10^4$ for the experimental parameters listed in Table~\ref{tab:OperatingParams}. The fractional frequency instability, as limited by noise of the optical detection, is
\begin{eqnarray}
\sigma_y^{(\mathrm{det})}(\tau) &=& \frac{1}{f_0} \frac{\Gamma_R}{\mathrm{SNR}} \frac{1}{\sqrt{\tau}} \nonumber \\
&=&1.0 \times 10^{-15}/\sqrt{\tau},
\end{eqnarray}
where $f_0=\omega_{21}/(2\pi)=6.8$\,GHz is the clock frequency and $\tau$ is the integration time. The SNR of the optical detection is almost two orders of magnitude larger than the SNR due to atom shot noise alone (SNR$\,\sim\sqrt{N_a}=10^3$ for $N_a=10^6$ atoms):
\begin{eqnarray}
\sigma_y^{(\mathrm{atoms})}(\tau) &=& \frac{1}{f_0} \frac{\Gamma_R}{\sqrt{N_a}} \frac{1}{\sqrt{\tau}} \nonumber \\
&=& 7.4 \times 10^{-14}/\sqrt{\tau}.
\end{eqnarray}
Aside from systematic effects and technical noise sources, clock performance will therefore be limited by the atom shot noise.


\section{Systematic Effects}
In this section, two systematic effects that impact the accuracy and long-term stability of the clock are considered. The first, AC Stark shifts, can be eliminated with a proper choice of laser parameters (e.g. intensities, EIT pulse duration, and common-mode laser detuning). The second set of shifts arise from the combined effects of magnetic field inhomogeneity and mean field shifts. The sensitivity of the clock frequency on relevant experimental parameters, including laser intensity, optical frequency, pulse timing, trap magnetic fields, and atom number is presented. The overall clock performance is calculated based on conservative estimates of drift in these parameters.

\subsection{AC Stark Shifts}
AC Stark shifts, or light shifts, are one of the most problematic sources of systematic error in both miniature and full-size atomic clocks. The term ``light shifts" refer to any effect that modifies the clock frequency through the atomic interaction with the laser fields (and other oscillating electromagnetic fields). The result is a clock whose accuracy ultimately depends on external experimental parameters instead of the inherent properties of the atomic species. In addition to accuracy, drift in these parameters can affect clock stability.

The advantage of pulsed EIT as a clock interrogation scheme is that a proper choice of Rabi frequencies (as determined by the laser intensities), EIT pulse duration, and common-mode laser detuning leads to a clock that is free of all light shifts. Our goal in this section is not to provide an accurate calculation that includes all sources of light shifts, which would be a computationally intensive task. Rather, we show that the shift-free configuration can be obtained for realistic operating parameters by just accounting for the the largest source of non-resonant light shifts. However, since the identification of shift-free parameters is best done experimentally, our numerically studies strongly suggest that an experimental approach would succeed in achieving a shift-free clock.

Light shifts can be divided into two categories. The first are shifts that arise from the resonant interaction of the two laser fields with the three-level atomic system (i.e. the idealized system represented in Fig.~\ref{fig:RbEnergyDiagram}). These shifts, which we label ``resonant", have been studied experimentally in atomic beams and have been found to arise when the initial EIT laser pulse is extinguished before the steady-state response has been reached. For the case of a closed atomic system with excited state decay rate $\Gamma$, the resonant light shift $\Delta \delta$ can be expressed mathematically as
\begin{eqnarray}
\label{eqn:ResonantShift}
\Delta \delta = -\frac{1}{T} \mathrm{tan}^{-1} \left( D_0 \frac{1}{e^{\alpha t_1}-1} \mathrm{sin}(\beta t_1)  \right),
\end{eqnarray}
where $D_0$ is the difference in the initial clock state populations~\cite{shahriar_dark-state-based_1997}. The Raman damping rate $\alpha$ and Raman dispersion $\beta$ are given by
\begin{eqnarray}
\label{eqn:ResonantShift2}
\alpha = \frac{1}{2} \frac{\Omega^2 \Gamma}{\Gamma^2 + 4 \Delta_0^2} \\
\beta = \frac{\Omega^2 \Delta_0}{\Gamma^2 + 4 \Delta_0^2},
\end{eqnarray}
where $\Omega$ is the Rabi frequency of both probe and coupling transitions (i.e. $\Omega_p=\Omega_c$) and the interrogation time $T$ is assumed to be much longer than the pulse duration $t_1$. The assumption of small common-mode detunings ($\Delta_0 \ll \Gamma$) yields $\alpha \approx \Omega^2/(2 \Gamma)$ and $\beta \approx \Omega^2\Delta_0/\Gamma^2$. The additional assumption that the argument of the arctangent is much less than 1 allows us to simplify Eqn.~\ref{eqn:ResonantShift} to
\begin{eqnarray}
\label{eqn:ResonantShiftSimple}
\Delta \delta = -\frac{D_0 \Delta_0}{T} \frac{\Omega^2}{\Gamma^2} \frac{t_1}{e^{\frac{\Omega^2 t_1}{2 \Gamma}}-1}.
\end{eqnarray}
Eqn.~\ref{eqn:ResonantShiftSimple} shows how the resonant shift depends on various experimental parameters. The shifts are inversely proportional to interrogation time $T$, reflecting the fact that the longer the atoms precess in the dark, the less the clock transition is perturbed by the lasers. The exponential term arises from the extinction of the first EIT pulse before the steady-state response is reached. This term is reduced by increasing $t_1$ or $\Omega$, which allows the system to reach equilibrium more quickly. In addition, shifts are reduced if the initial clock state populations are closer to their steady-state values. As a result, the shift is proportional to $D_0$. Last, the resonant light shift can be made positive or negative via the linear dependence on $\Delta_0$.

\begin{figure}
\centering
\includegraphics[width=3.4in]{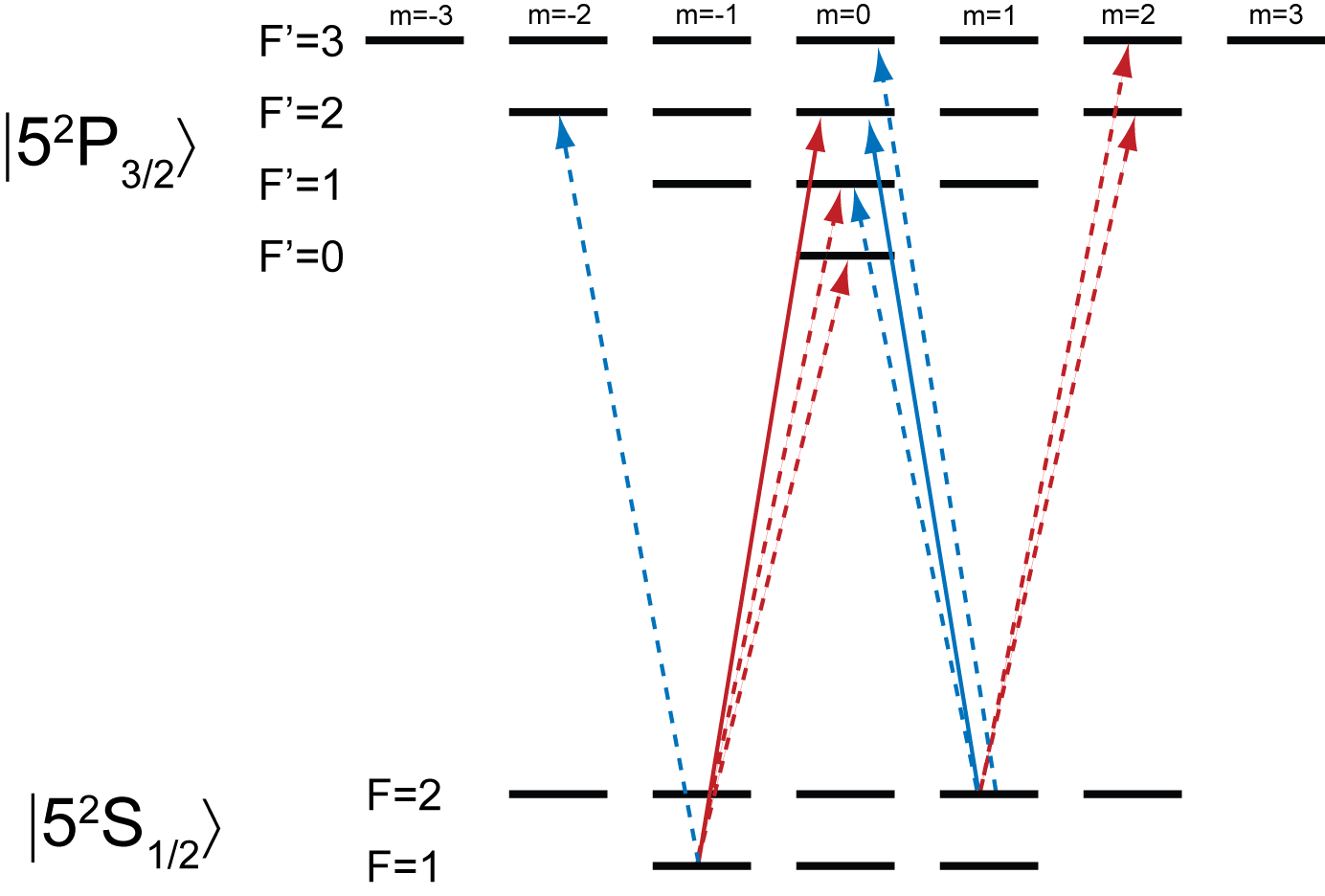}
\caption{Energy diagram of the $5^2S_{1/2}$ ground state and $5^2P_{3/2}$ excited state hyperfine manifolds in $^{87}$Rb. Resonant (solid line) and non-resonant (dashed line) transitions allowed by the probe (coupling) laser are shown in red (blue). Only transitions that couple to the clock states in the ground state manifold are considered.}
\label{fig:StarkShiftDiagram}
\end{figure}
Additional AC Stark shifts arise from the presence of other atomic states. We label these shifts ``non-resonant" since the lasers that couple these transitions are detuned by many linewidths. In general, these shifts scale inversely with laser detuning, so the dominant contribution will come from other nearby magnetic sublevels in the $5^2S_{1/2}$ ground state and $5^2P_{3/2}$ excited state hyperfine manifolds (see Fig.~\ref{fig:StarkShiftDiagram}). Although there are 24 sublevels in total, only three of the eight sublevels in the ground state can be magnetically trapped; contributions from the five other untrappable ground-state sublevels are ignored. Furthermore, the $|F=2,m_F=2\rangle$ ground state sublevel, which can be magnetically trapped, cannot be populated via spontaneous decay of an $m'=0$ excited level, leaving the two clock states as the only ground state sublevels that need consideration. Of the sixteen excited state sublevels, only six can be non-reson-antly coupled to the clock states with circularly polarized light.

The excited non-resonant sublevel that produces the largest AC Stark shift is $|4\rangle=|F'=1,m_{F'}=0\rangle$. There are two reasons for this. First, the state $|4\rangle$ is the only non-resonant excited sublevel that couples to both laser fields. Second, non-resonant transitions to excited sublevels with $|m'|=2$ are coupled with a laser detuning of $\sim6.8$\,GHz, while the detunings for transitions to $m'=0$ sublevels are given by the smaller 100 to 300\,MHz splittings of the excited $5^2P_{3/2}$ manifold. In fact, for our choice of state $|3\rangle$, the detuning of $\sim157\,$MHz for state $|4\rangle$ is smaller than that of all other $m'=0$ non-resonant sublevels.

To cancel light shifts in the clock, a resonant light shift that is equal in magnitude, but opposite in sign, to the sum of all non-resonant light shifts is intentionally introduced; the resulting resonant and non-resonant light shifts sum to zero. To demonstrate this cancelation numerically, the density matrix in Eqn.~\ref{rho_and_sigma} was enlarged to a 4$\times$4 matrix for the inclusion of state $|4\rangle$ in the system dynamics. The expanded density matrix equations were solved and the integrated optical phase shift was calculated as a function of $\delta$ over the center part of the central Ramsey fringe. The clock frequency is given by the value of $\delta$ at which the fringe signal equals 0, which was found numerically for different parameters.

Several assumptions were made in solving the 4-level system. First, the slaved-state approximation was made by setting the time-derivative of the coherence between the two excited states equal to 0. This is equivalent to assuming that the population of state $|4\rangle$ is negligible. Second, the resonant light shifts in Eqn.~\ref{eqn:ResonantShiftSimple} depend only on the duration $t_1$ of the first EIT pulse. Non-resonant light shifts, on the other hand, arise for any time the laser fields are present, and are therefore affected by the duration of both pulses. To simplify the calculations, the first and second pulse durations are assumed to be equal, or $t_1=t_2$. Third, the Rabi frequencies are assumed to both equal a fixed value $\Omega_c=\Omega_p=(2\pi)150\,$kHz, and the initial atomic population is assumed to be entirely in state $|1\rangle$ so that $D_0=1$. As a result, the resonant light shift in Eqn.~\ref{eqn:ResonantShiftSimple} is modified by changing the pulse durations and common-mode detuning $\Delta_0$. Fourth, both resonant and non-resonant light shifts scale inversely with interrogation time $T$ in the limit $T\gg t_1$; the value $T=1\,$s was used for all calculations.

\begin{figure}
\centering
\includegraphics[width=3.2in]{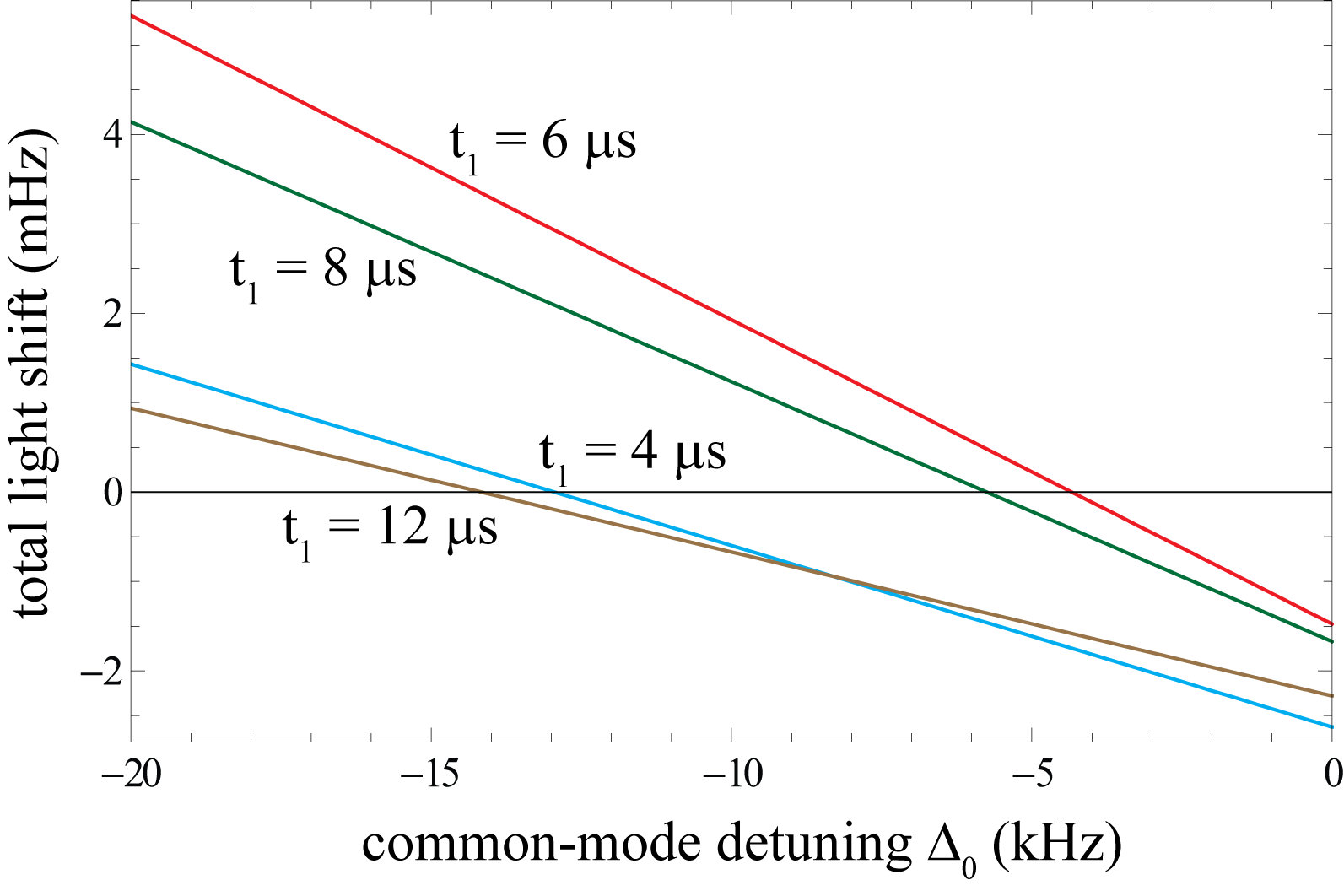}
\caption{The total light shift as a function of common-mode detuning $\Delta_0$ for different values of the pulse duration $t_1$.}
\label{fig:StarkShiftCancellation1}
\end{figure}

The dependence of the clock frequency on pulse duration is presented in Fig.~\ref{fig:StarkShiftCancellation1}, which shows the total light shift scaling linearly with $\Delta_0$. For values of $t_1 = t_2$ between 4 and 12\,$\mu$s, the total light shift can be eliminated for detunings $\Delta_0$ between -4 and -14\,kHz. The steepest slope of -0.34\,mHz/(kHz $T$) was found for $t_1\sim6\,\mu$s. This worst-case dependency of the clock performance on fluctuations in $\Delta_0$ corresponds to a fractional frequency shift of $-5.0\times 10^{-14}$.

\begin{figure}
\centering
\includegraphics[width=3.2in]{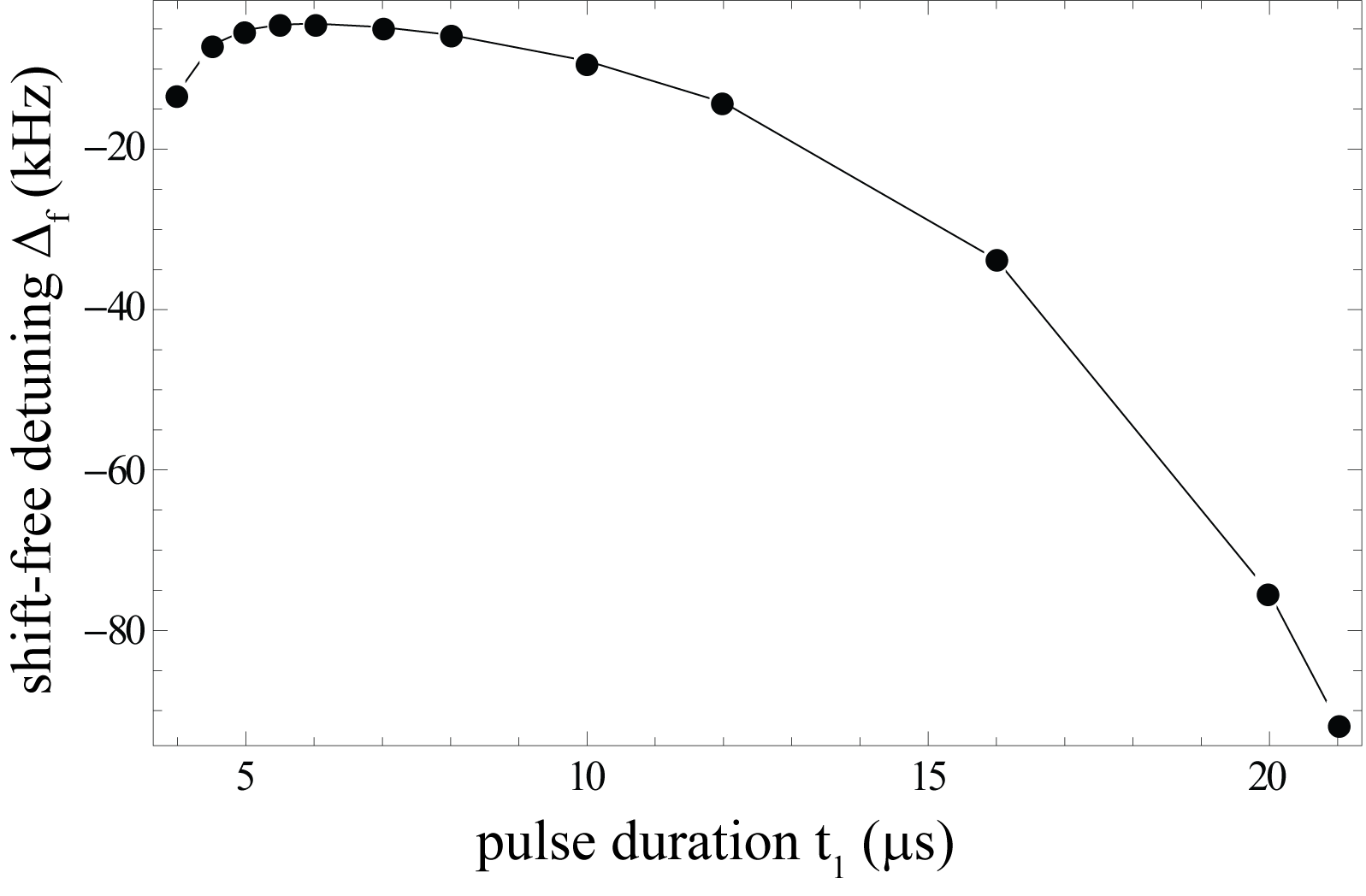}
\caption{Shift-free value $\Delta_f$ of the common-mode detuning as a function of pulse duration $t_1$.}
\label{fig:MagicDetuning}
\end{figure}

The common-mode detuning $\Delta_f$ at which light shifts cancel is plotted as a function of pulse duration in Fig.~\ref{fig:MagicDetuning}. The maximum value of $\Delta_f$ occurs near $6\,\mu$s. At this point, the clock will be least sensitive to fluctuations in pulse duration. However, this point also corresponds to the greatest sensitivity to fluctuations in $\Delta_0$. Therefore, pulse durations different from 6\,$\mu$s should also be considered, even though this will increase sensitivity to timing jitter.

Clock shifts arising from fluctuations in the laser parameters were estimated by numerically calculating the total light shift for small changes of these parameters. Table~\ref{tab:ParameterSensitivities} summarizes the results for $t_1=6\,\mu$s and $t_1=20\,\mu$s. The units are normalized to typical uncertainties of the corresponding parameters (i.e. a 1\,kHz uncertainty in $\Delta_0$, a 0.1\% shift in $t_1$, etc.). Even though we assume the first and second pulses are equal in duration, timing jitter affects light shifts differently for the two pulses. Therefore, we consider the effects of timing jitter for each pulse separately. Changes in the Rabi frequencies arise from laser intensity variations and are expected to be the largest source of laser-induced inaccuracy for $t_1=20\,\mu$s. For $t_1=6\,\mu$s, the dominant sensitivity is due to $\Delta_0$, whose fluctuations depend on how stable the optical frequencies are. For both values of $t_1$, the clock is least sensitive to timing jitter in the laser pulses. The overall performance is slightly better for $t_1=20\,\mu$s due to the reduced sensitivity to $\Delta_0$. In this case, all inaccuracies due to laser-induced effects are less than $2\times10^{-14}$. Active stabilization of the laser intensity could reduce this even further.

The dependency of the clock frequency on the Rabi frequencies, even under light-shift-free conditions, implies that the $\Delta_f$ cannot be experimentally identified by varying probe and coupling laser intensities. Instead, $\Delta_f$ can be found by measuring the clock frequency as a function of $\Delta_0$ for different values of $T$. Since both resonant and non-resonant light shifts scale inversely with $T$, the clock frequency varies linearly with $\Delta_0$ with a slope proportional to $1/T$. When plotting the shifts for all values of $T$ on the same graph, the lines corresponding to different values of $T$ will intersect at the shift-free detuning $\Delta_f$ (see Fig.~3c in Reference~\cite{zanon-willette_cancellation_2006}).

\begin{table}
\centering
\begin{tabular}{c|c|c|c}
\hline\noalign{\smallskip}
Parameter & $t_1=6\,\mu$s  &  $t_1=20\,\mu$s & units \\
\noalign{\smallskip}\hline\noalign{\smallskip}
$\Delta_0$  & $-5.0 \times 10^{-14}$ & $-7.0 \times 10^{-15}$  & (kHz $T$)$^{-1}$   \\
$t_1$       & $+5.9 \times 10^{-17}$ & $-1.7 \times 10^{-15}$  & (0.1\% $T$)$^{-1}$ \\
$t_2$       & $-6.0 \times 10^{-17}$ & $-4.2 \times 10^{-16}$  & (0.1\% $T$)$^{-1}$ \\
$\Omega_p$  & $+2.6 \times 10^{-16}$ & $-8.2 \times 10^{-15}$  & (0.5\% $T$)$^{-1}$ \\
$\Omega_c$  & $+5.4 \times 10^{-16}$ & $+1.3 \times 10^{-14}$  & (0.5\% $T$)$^{-1}$ \\
\noalign{\smallskip}\hline
\end{tabular}
\caption{Light shifts arising from uncertainties in the common-mode detuning $\Delta_0$, pulse durations $t_1$ and $t_2$, and Rabi frequencies $\Omega_c$ and $\Omega_p$. Values are obtained
for $t_1=6\,\mu$s and $t_1=20\,\mu$s assuming $\Omega_c=\Omega_p=(2\pi)150\,$kHz, $t_1=t_2$, and $\Delta_0=\Delta_f$. Shifts are inversely proportional to $T$. Units are expressed in terms of typical uncertainties of the corresponding parameters.}
\label{tab:ParameterSensitivities}
\end{table}


\subsection{Dephasing and Coherence Times}
After the atoms have been pumped into a coherent superposition state, two effects lead to dephasing and systematic frequency shifts: differential Zeeman shifts and mean-field shifts. These two effects generate a spread in clock frequencies $\Delta \nu$ that cause the atoms to precess at different rates, washing out the Ramsey fringes. These two effects can be made to largely cancel each other for a proper choice of trap magnetic fields.

Differential Zeeman shifts arise from the inhomogeneous magnetic field of the trap. In a harmonic trap, hotter atoms with greater kinetic energy are able to travel farther from the trap center, where they are exposed to larger magnetic fields and therefore greater Zeeman shifts. The magnetic field of the trap can be written
\begin{eqnarray}\label{eqn:trapBfield}
B = B_m + \alpha_t \left(x^2 + y^2\right) + \alpha_l z^2,
\end{eqnarray}
where $B_m$ is the minimum value of the field at the center of the trap; $x$, $y$, and $z$ are the trap's spatial coordinates; and $\alpha_t$ and $\alpha_l$ are the field curvatures in the transverse and longitudinal directions, respectively. In the presence of a weak magnetic field, the potential energy $U(x,y,z)$ of an atom scales linearly with the field:
\begin{eqnarray}\label{eqn:HarmonicTrapPotential}
&& \hspace{-0.5in} U(x,y,z) = g_F\,m_F\,\mu_B\,[B_m \nonumber \\
&& {} + \alpha_t \left(x^2+y^2\right) + \alpha_l z^2],
\end{eqnarray}
where $g_F$ and $m_F$ are the $g$-factor and magnetic quantum number of the atomic state, $\mu_B$ is the Bohr magneton, and $m$ is the mass of the atom. The product $g_F\,m_F$ is equal to $1/2$ for both clock states.

The Breit-Rabi formula predicts that the linear Zeeman shifts of the two clock states are equal at $B_0\approx3.23$\,G. Near $B_0$, the second-order Zeeman shift of the clock transition can be expressed
\begin{eqnarray}\label{eqn:QuadraticZeemanShift}
\Delta \nu_z= b(B-B_0)^2 - \nu_m,
\end{eqnarray}
where $b\,=\,431$\,Hz/G$^2$ and $\nu_m\,\approx\;$4497\,Hz. Plugging \linebreak Eqn.~\ref{eqn:HarmonicTrapPotential} into Eqn.~\ref{eqn:QuadraticZeemanShift} gives the position-dependent Zeeman shift $\Delta \nu_z(x,y,z)$ of the clock transition
\begin{eqnarray}
\lefteqn{\Delta \nu_z(x,y,z) =} \nonumber \\
&& b\left[B_m-B_0 + \alpha_t (x^2+y^2) + \alpha_l z^2\right]^2-\nu_m.
\end{eqnarray}
The Zeeman shift varies as the fourth power of position. More importantly, the sign of $B_m-B_0$ can be changed to make the curvature at the trap center positive ($B_m>B_0$), negative ($B_m<B_0$), or flat ($B_m=B_0$).

\begin{table*}
\centering
\begin{tabular}{c|c|c|c|c}
\hline\noalign{\smallskip}
Parameter & Sensitivity  & Parameter Uncertainty & Frequency Uncertainty (mHz) & Fractional Frequency Uncertainty \\
\noalign{\smallskip}\hline\noalign{\smallskip}
$B_m$   & -35 Hz/G           & $10^{-5}=30\,\mu$G       &  -1.1         & 1.5$\times10^{-13}$    \\
$T_a$   & 1 Hz/$\mu$K        & $10^{-5}=5\,$pK          &   0.005       & 7.4$\times10^{-16}$    \\
$N$     & -5 $\mu$Hz/atom    & $\sqrt{N}=10^3$          &  -5           & 7.4$\times10^{-13}$    \\
$f$     & 0.3 mHz/\%         & 0.5\%                    &   0.15        & 2.2$\times10^{-14}$    \\
\noalign{\smallskip}\hline
\end{tabular}
\caption{Differential Zeeman shifts and mean-field shifts arising from uncertainties in the trap bias field $B_m$, atom temperature $T_a$, atom number $N$, and population partition $f$. Typical uncertainties in the parameters, and the corresponding shifts in the clock frequency are listed. Sensitivities in the second column are taken from reference~\cite{rosenbusch_magnetically_2009}.}
\label{tab:ZeemanShiftTable}
\end{table*}

For cold, non-condensed atoms in the s-wave regime, atom-atom collisions generate a frequency shift that depends on scattering length and atomic density. For the clock transition, this mean-field shift can be expressed
\begin{eqnarray}\label{eqn:ColdCollisionalShift}
\Delta \nu_c &=& \frac{2 \hbar}{m}\left[\alpha_{12} a_{12} (n_1-n_2) + \alpha_{22} a_{22} n_2 - \alpha_{11} a_{11} n_1\right] \nonumber \\
&=& \frac{\hbar}{m} n [\alpha_{22} a_{22}-\alpha_{11} a_{11} \nonumber \\
&& {} + f(2 \alpha_{12} a_{12} - \alpha_{11} a_{11} - \alpha_{22} a_{22}) ],
\end{eqnarray}
where $m$ is the atomic mass, $n_1$ and $n_2$ are the atomic densities for states $|1\rangle$ and $|2\rangle$, $n=n_1+n_2$ is the total atomic density, and the partition $f=(n_1-n_2)/n$ is the fractional change in population densities due to unequal superpositions. We assume the atoms are pumped into equal superposition states (i.e. $f=0$). The s-wave scattering lengths are represented by $a_{ij}$ and the zero-distance two-atom correlation functions are given by $\alpha_{ij}$. For non-condensed atoms, $\alpha_{11}=\alpha_{22}=\alpha_{12}=2$. Using the measured scattering lengths of $a_{22}=95.47a_0$, $a_{11}=100.44 a_0$, and $a_{12}=98.09a_0$~\cite{van_kempen_interisotope_2002}, where $a_0$ is the Bohr radius, Eqn.~\ref{eqn:ColdCollisionalShift} can be simplified to
\begin{eqnarray}
\Delta \nu_c (x,y,z) = -4.97 \frac{2 \hbar}{m} a_0 n(x,y,z),
\end{eqnarray}
which shows that the shift is both negative and proportional to atom density $n(x,y,z)$. Due to the Gaussian density profile of the ensemble, atoms at the center of the trap experience a more negative shift than atoms at the edge of the trap.

The mean-field shift can be mostly canceled by the differential Zeeman shift when $B_m$ is chosen to be slightly less than $B_0$. In this case, the Zeeman shift $\Delta \nu_z$ has positive curvature at the trap center, in contrast to the negative curvature of the collisional shift. Taken together, the combined shift is nearly flat across the spatial extent of the trap, thus reducing the spread of clock frequencies and increasing coherence times.

A theoretical framework has been developed in reference~\cite{rosenbusch_magnetically_2009} for better estimating dephasing times of Rb atoms in a magnetic trap. In this work, the total shift $\Delta \nu_z+\Delta \nu_c$ was either spatially averaged over a Maxwell-Boltzmann distribution of atoms of finite temperature or averaged over the motion of the atoms in the trap. The resulting average shift was used to compute the variance $\sigma^2$, from which the dephasing time is given by $(\sqrt{2}\pi\sigma)^{-1}$. Calculations showed good agreement with previous experimental data~\cite{treutlein_coherence_2004,harber_effect_2002}.

This framework was used to both minimize the line spread of the clock transition and to increase the dephasing time by varying atom temperature, bias magnetic field, peak atom density, atom number, population partition, and trap frequencies~\cite{rosenbusch_magnetically_2009}. Dephasing times as long as 20\,s, corresponding to a line spread of 11 mHz, were obtained for experimental parameters similar to those used in references~\cite{treutlein_coherence_2004} and~\cite{harber_effect_2002}: a temperature $T_a=500\,$nK, peak densities of $\sim2.0\times 10^{12}$/cm$^3$, a total atom number $N=10^5$, a population partition $f=0$, axial trap frequencies between 1\,Hz and 10\,Hz, and transverse trap frequencies between 300\,Hz and 1\,kHz.

The effects of fluctuations in magnetic field, temperature, atom number, and partition on clock performance are summarized in Table~\ref{tab:ZeemanShiftTable}, with values for the sensitivities coming from reference~\cite{rosenbusch_magnetically_2009}. The largest fractional frequency sensitivity is due to fluctuations in atom number $N$. Unlike drift in the other parameters, fluctuations in atom number have a white noise spectrum, and therefore this source of clock instability will improve as $1/\sqrt{\tau}$ for repeated clock cycles (assuming each clock cycle starts with the same nominal number of atoms). Although this systematic effect will slightly degrade stability, it will not contribute to the noise floor that will ultimately limit the long-term stability of the clock.

The stability of the trap bias field $B_m$ is limited by the stability of the electrical current flowing through the atom chip; a $10^{-5}$ relative stability can be readily achieved. Since the final atom temperature is determined by the RF evaporation frequency and bias field $B_m$, the temperature should have the same relative stability as $B_m$. One source of drift in the population partition $f$ is fluctuations in the Rabi frequencies of the two optical transitions. For probe and coupling laser intensities stable to 1\%, the Rabi frequencies, and therefore the partition $f$, will be stable to 0.5\%. The effects of fluctuations in the trap frequencies are negligible~\cite{rosenbusch_magnetically_2009}.

Given that the magnetic trap is expected to be the largest source of systematic error, we briefly consider a miniature clock that confines atoms optically (e.g. an optical dipole trap). This approach could require several Watts of laser power, however this is not necessarily a larger power drain than what is already needed for magnetically trapping on a microchip. Similar to a fountain clock, this configuration could use the $m=0$ ground state sublevels for the clock transition, helping to reduce sensitivity to stray magnetic fields. Due to the lack of a magic wavelength for alkali atomic clocks based on optical dipole traps, an additional light shift that depends on the intensity and wavelength of the trapping laser will arise~\cite{rosenbusch_ac_2009,beloy_micromagic_2009,flambaum_magic_2008}. Although further calculations for the case of Rb are needed, this light shift could in principle be eliminated either using the pulsed EIT scheme presented here~\cite{zanon-willette_cancellation_2006} or with the addition of an external magnetic field~\cite{flambaum_magic_2008,Derevianko2009,Lundblad2009}. The result would be an atomic clock free of both light shifts and Zeeman shifts.


\section{Description of Proposed Apparatus}
\label{section:DescriptionOfApparatus}
In this section, we present a overview of how recent technological developments in ultracold atom technology can be adopted for the creation of a miniature, transportable atomic clock. There are four critical elements to the proposed system: the miniature ultrahigh-vacuum (UHV) cell with integrated atom microchip; magnets; the laser system, optics, and detection system; and control electronics with a low phase noise microwave oscillator. Our focus here is primarily on the UHV cell and laser system. The optical detection system has already been discussed above (see Fig.~\ref{fig:HeterodyneDetection}). Miniaturization of control electronics and low phase noise microwave oscillators has already been developed for CSACs, and our proposal relies on the same technology.

\begin{figure*}
\centering
\includegraphics[width=6in]{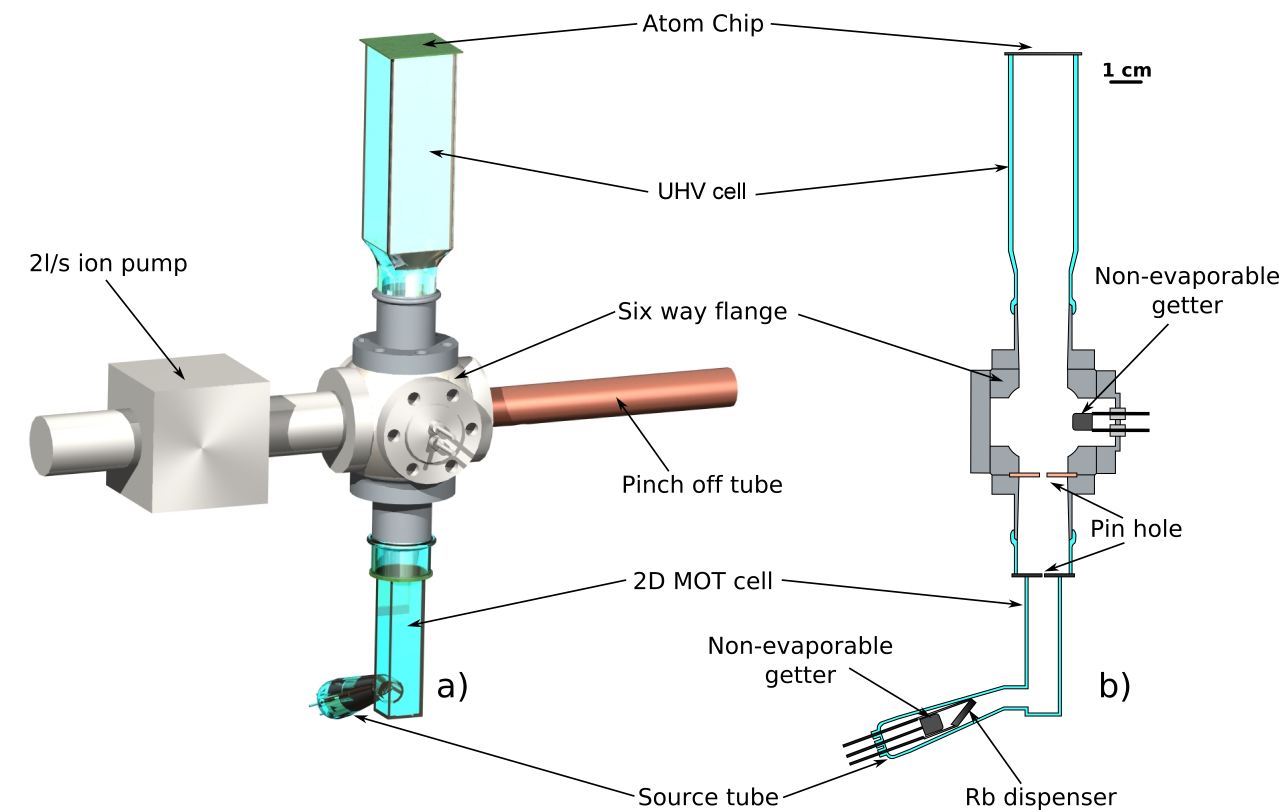}
\caption{Schematic of the double MOT vacuum cell system.}
\label{fig:TwoChamberCell}
\end{figure*}

The clock will be based on a double MOT vacuum cell system (see Fig.~\ref{fig:TwoChamberCell}) that has been used for several cold-atom experiments and applications in our lab at the University of Colorado at Boulder. The double MOT configuration offers a substantial advantage in providing high atom flux, fast loading times, and long MOT and magnetic trap lifetimes. With an internal volume of 30\, cm$^3$, this cell has a vacuum level better than $5\times10^{-10}$\,Torr and has been used to create Bose-Einstein condensates near the surface of an integrated atom microchip with a duty cycle as fast as 3\,s~\cite{Farkas2009}. The microchip is anodically bonded onto the end of one of the chambers, where it forms a wall of the cell. Current is passed into the vacuum through UHV-compatible electrical vias fabricated on the chip. Low vacuum levels are maintained by a 2\,l/s ion pump and non-evaporable getters. Cell construction utilizes silicon-to-glass anodic bonding and standard metal, glass, and glass-to-metal construction methods. These construction techniques enable high-temperature bake-outs up to 300$^{\circ}$C, which is essential for good UHV performance.

The cell consists of two chambers joined by a 0.75\,mm aperture that permits differential pumping between the two chambers. In the lower chamber, a 2D$^+$ MOT produces a beam of cold atoms, which feeds the upper cell, where the atoms are captured in a 3D MOT. The 2D$^+$ MOT produces a captured atom flux of 0.5 to 1 billion atoms/s with less than 120\,mW of total optical power. The 2D$^+$ MOT chamber is operated near $10^{-7}$\,Torr of rubidium pressure for high capture efficiency, while the 3D MOT chamber is maintained at a much lower pressure for long trap lifetimes. Typical trap lifetimes are 10 to 20\,s for the 3D MOT, and 4 to 7\,s for a Ioffe-Pritchard magnetic trap.

After capture in the 3D MOT, the magnetic fields are turned off and a few ms of polarization gradient cooling (PGC) is applied to bring the atom temperature below 50\,$\mu$K. The atoms are then optically pumped into a single, trappable magnetic sublevel. External magnetic field coils are used to transport atoms a distance of $\sim1\,$\,cm in 300 to 500\,ms, leaving the cold atomic cloud within 0.5\,mm of the vacuum side of the atom microchip. Atoms are transferred onto the chip by simultaneously ramping up current flowing on the chip and ramping down current flowing through the external field coils. The chip conductor forms a ``Z" configuration that, in conjunction with a pair of external bias fields parallel to the chip surface, creates a Ioffe-Pritchard trap approximately 100\,$\mu$m from the chip surface. Evaporative cooling with an RF antenna reduces the atom temperature below 1\,$\mu$K.

There are several benefits to using atom microchips. First, the significantly higher trap frequencies that can be obtained (1-2 kHz in the transverse direction) compared to macroscopic magnetic traps speeds up rethermalization during evaporative cooling, helping to shorten the duration of this stage. Second, less RF power can be used during evaporation by placing the RF antenna on the back-side of the chip, where it is less than 0.5\,mm away from the atoms. In addition, the ability to generate very stable electrical currents may lead to trap fields that are more stable than what can be obtained with permanent magnets, which have at best a temperature coefficient of 0.03\%/$^{\circ}$C.

One drawback to using an atom microchip is power consumption. The chip trap must be on during evaporative cooling and clock interrogation, for a total of several seconds. Assuming a wire resistance of 2\,$\mathrm{\Omega}$ and a steady-state current of 1.5\,A, the resulting power dissipation is more than 4\,W (ignoring the additional power needed to generate external bias fields). One way to mitigate this loss is to use a planar configuration of permanent magnets on the chip as the final trap~\cite{barb_creating_2005}. Chip wires can then be used for the much shorter stage of transferring atoms to the chip.

Distributed feedback (DFB) diode lasers can be used to create a compact laser system that, compared to external cavity diode lasers, is significantly more robust towards mechanical vibrations. In addition to the one or two lasers needed for EIT, up to two additional lasers are needed for the MOT: one for the $|5S_{1/2},F=2\rangle \rightarrow |5P_{3/2},F'=3\rangle$ cooling transition and the other for the $|5S_{1/2},F=1\rangle \rightarrow |5P_{3/2},F'=2\rangle$ repumping transition. MOTs of 1 to 2 billion atoms have been achieved in the two-chamber cell of Fig.~\ref{fig:TwoChamberCell} using 120\,mW of combined cooling and repump power. However, to obtain 1 million atoms in a magnetic trap after evaporative cooling, we estimate only 100 to 200 million atoms will be needed in the 3D MOT. As a result, we expect only 20 to 30\,mW of laser light will be necessary. By tuning the power and frequencies of the cooling and repump lasers, these lasers could also serve as the EIT probe and coupling lasers. This option would reduce volume and power consumption by minimizing the number of lasers in the system.


\section{Timing and Power Consumption}

Table~\ref{tab:TimingSequence} lists timing and power consumption estimates for one operational cycle of the clock. The total duration is between 3 and 6\,s, and the clock instabilities calculated in Section~\ref{sec:SNR} must be increased by the square-root of this time. The dominant power drains in the physics package are the RF signal used for evaporative cooling ($\sim1\,$W); the power dissipated by the chip wires ($\sim5\,$W) during evaporative cooling and clock interrogation; and the power used to drive external magnetic field coils.

\begin{table}
\centering
\begin{tabular}{l|cc}
\hline\noalign{\smallskip}
Stage & Duration (ms) & Power (W) \\
\noalign{\smallskip}\hline\noalign{\smallskip}
MOT Cooling           & 1000            & 0.1    \\
PGC                   & 5-10            & 0.1    \\
Transfer to Chip      & 300-500         & 5      \\
Evaporative Cooling   & 1000-4000       & 6      \\
First EIT Pulse       & $\sim$0.01      & 5      \\
Interrogation         & 1000            & 5      \\
Second EIT Pulse      & $\sim$0.01      & 5      \\
\noalign{\smallskip}\hline
\end{tabular}
\caption{Timing and power consumption estimates for one operational cycle of the proposed clock.}
\label{tab:TimingSequence}
\end{table}

Operation of the cold-atom clock in conjunction with a CSAC provides a means of duty-cycled operation to reduce overall power consumption. For example, the cold-atom apparatus can be operated once every hundred seconds, providing periodic frequency corrections to a continuously-running CSAC. Since the power duty cycle is reduced to 1\%, the average power consumed by the cold-atom physics package is reduced by two orders of magnitude.


\section{Conclusion and Outlook}

We have investigated the feasibility of a transportable, miniature atom clock based on laser cooling and magnetic trapping of $^{87}$Rb atoms near the surface of an atom microchip. A Ramsey sequence based on pulsed EIT, in conjunction with a heterodyne optical detection approach that can detect optical powers as low as 100 pW at the shot noise level, can achieve detection almost two orders of magnitude more sensitive than performance limited by the shot-noise of 10$^6$ atoms. By constructing the clock around a compact, UHV double MOT cell, interrogation times up to 1\,s should be attainable, resulting in Ramsey fringes as narrow as 0.5\,Hz. Through the use of an integrated atom microchip, this compact geometry both improves clock stability by increasing duty cycle and reduces power consumption by minimizing the volume of external magnetic field coils.

As a clock interrogation scheme, pulsed EIT permits implementation of a novel light-shift cancelation scheme. Here, resonant light shifts arising from the laser coupling of the clock states to a common excited state are intentionally introduced in order to cancel non-resonant light shifts arising from all other atomic states. The density matrix equations were solved with the inclusion of a fourth atomic state that generates the largest non-resonant light shift; cancelation of light shifts was numerically demonstrated for realistic values of the EIT pulse duration and common-mode laser detuning. Estimating the effects of fluctuations in laser intensities, pulse timings, and common-mode detuning on clock performance, residual light shifts below $2\times10^{-14}$ should be achievable assuming reasonable values for drift and uncertainty in these parameters.

Clock stability will be limited by shot-noise of the atoms, which affect both the SNR and the combined effects of differential Zeeman and mean field shifts. Additionally, a slight reduction in stability arises from a duty cycle less than 100\%. Taking all these effects into account, a fractional frequency instability of $1\times10^{-13}/\sqrt{\tau}$ should be achievable. The accuracy of the clock should be limited by differential Zeeman and mean field shifts to better than $1\times10^{-13}$.

\section*{Acknowledgements}

This work was supported in part by the National Science Foundation through a Physics Frontier Center \linebreak (PHY0551010).


%

\end{document}